\documentclass[longauth]{aa}  

\usepackage{graphicx}
\usepackage{tabularx}
\usepackage{booktabs}
\usepackage{multirow}
\usepackage{txfonts}
\usepackage{cleveref}
\usepackage{colortbl} % color tables

\crefname{equation}{Eq.}{Eqs.}
\Crefname{equation}{Equation}{Equations}
\crefname{figure}{Fig.}{Figs.}
\Crefname{figure}{Figure}{Figures}
\crefname{table}{Table}{Tables}
\Crefname{table}{Table}{Tables}
\crefname{section}{Section}{Sections}
\Crefname{section}{Section}{Sections}

\newcolumntype{L}[1]{>{\raggedright\arraybackslash}p{#1}}
\newcolumntype{R}[1]{>{\raggedleft\arraybackslash}p{#1}}

\begin{document} 

%TITLE%PAGE%%%%%%%%%%%%%%%%%%%%%%%%%%%%%%%%%%%%%%%%%%%%%%%%%%%%%%%%%%%%%%%%%%%%

%_TITLE________________________________________________________________________
\title{Polarization angle swings in blazars: The case of \object{3C 279}}

% \thanks{The measured and processed
%   optical polarization and R-band photometry data are available in electronic 
%   form at the CDS via anonymous ftp to cdsarc.u-strasbg.fr (130.79.128.5)
%  or via \url{http://cdsweb.u-strasbg.fr/cgi-bin/qcat?J/A+A/}}

%_AUTHORS______________________________________________________________________
\author{
  S.~Kiehlmann\inst{1,2,3}\fnmsep\thanks{\email{sebastian.kiehlmann@aalto.fi}}
  \and T.~Savolainen\inst{1,2,3}
  \and S.~G.~Jorstad\inst{4,5}
  \and K.~V.~Sokolovsky\inst{6,7,8}
  \and F.~K.~Schinzel\inst{9}
  \and A.~P.~Marscher\inst{4}
  \and V.~M.~Larionov\inst{5,10}
  \and I.~Agudo\inst{11}
  \and H.~Akitaya\inst{12}
  \and E.~Ben\'itez\inst{13}
  \and A.~Berdyugin\inst{14}
  \and D.~A.~Blinov\inst{15,16,5}
  \and N.~G.~Bochkarev\inst{7}
  \and G.~A.~Borman\inst{17}
  \and A.~N.~Burenkov\inst{18}
  \and C.~Casadio\inst{11}
  \and V.~T.~Doroshenko\inst{19}
  \and N.~V.~Efimova\inst{10}
  \and Y.~Fukazawa\inst{20}
  \and J.~L.~G\'omez\inst{11}
  \and T.~S.~Grishina\inst{5}
  \and V.~A.~Hagen-Thorn\inst{5}
  \and J.~Heidt\inst{21}
  \and D.~Hiriart\inst{22}
  \and R.~Itoh\inst{20}
  \and M.~Joshi\inst{4}
  \and K.~S.~Kawabata\inst{12}
  \and G.~N.~Kimeridze\inst{23}
  \and E.~N.~Kopatskaya\inst{5}
  \and I.~V.~Korobtsev\inst{24}
  \and T.~Krajci\inst{25}
  \and O.~M.~Kurtanidze\inst{23,21, 26}
  \and S.~O.~Kurtanidze\inst{23}
  \and E.~G.~Larionova\inst{5}
  \and L.~V.~Larionova\inst{5}
  \and E.~Lindfors\inst{14}
  \and J.~M.~L\'opez\inst{22}
  \and I.~M.~McHardy\inst{27}
  \and S.~N.~Molina\inst{9}
  \and Y.~Moritani\inst{28}
  \and D.~A.~Morozova\inst{5}
  \and S.V.~Nazarov\inst{18}
  \and M.~G.~Nikolashvili\inst{23}
  \and K.~Nilsson\inst{29}
  \and N.~G.~Pulatova\inst{17}
  \and R.~Reinthal\inst{14}
  \and A.~Sadun\inst{30}
  \and M.~Sasada\inst{4}
  \and S.~S.~Savchenko\inst{5}
  \and S.~G.~Sergeev\inst{17}
  \and L.~A.~Sigua\inst{23}
  \and P.~S.~Smith\inst{31}
  \and M.~Sorcia\inst{13}
  \and O.~I.~Spiridonova\inst{18}
  \and K.~Takaki\inst{20}
  \and L.~O.~Takalo\inst{14}
  \and B.~Taylor\inst{4,32}
  \and I.~S.~Troitsky\inst{5}
  \and M.~Uemura\inst{12}
  \and L.~S.~Ugolkova\inst{7}
  \and T.~Ui\inst{20}
  \and M.~Yoshida\inst{12}
  \and J.~A.~Zensus\inst{3}
  \and V.~E.~Zhdanova\inst{18}
  }

%_AFFILIATIONS_________________________________________________________________
\institute{
  Aalto University Mets\"ahovi Radio Observatory, Mets\"ahovintie 114, 02540 Kylm\"al\"a, Finland
  \and Aalto University Department of Radio Science and Engineering, PL 13000, 00076 Aalto, Finland
  \and Max-Planck-Institut f\"ur Radioastronomie, Auf dem H\"ugel, 69, D-53121, Bonn, Germany
  \and Institute for Astrophysical Research, Boston University, 725 Commonwealth Avenue, Boston, MA 02215, USA
  \and Astronomical Institute, St. Petersburg State University, Universitetskij Pr. 28, Petrodvorets, 198504 St. Petersburg, Russia
  \and Astro Space Center of Lebedev Physical Institute, Profsoyuznaya 84/32, 117997, Moscow, Russia
  \and Sternberg Astronomical Institute, M.V.Lomonosov Moscow State University, Universiteskij prosp. 13, Moscow 119991, Russia
  \and Institute of Astronomy, Astrophysics, Space Applications and Remote Sensing, National Observatory of Athens\\ Vas.~Pavlou~\&~I.~Metaxa, GR-15~236 Penteli, Greece
  \and Department of Physics and Astronomy, University of New Mexico, Albuquerque NM, 87131, USA
  \and Main (Pulkovo) Astronomical Observatory of RAS, Pulkovskoye shosse, 60, 196140, St. Petersburg, Russia
  \and Instituto de Astrof\'{\i}sica de Andaluc\'{\i}a, CSIC, Apartado 3004, 18080, Granada, Spain
  \and Hiroshima Astrophysical Science Center, Hiroshima University, Higashi-Hiroshima, Hiroshima 739-8526, Japan
  \and Instituto de Astronom\'ia, Universidad Nacional Aut\'onoma de M\'exico, 04510 M\'exico DF, M\'exico
  \and Tuorla Observatory, Department of Physics and Astronomy, University of Turku, V\"ais\"al\"antie 20, 21500, Piikki\"o, Finland
  \and Foundation for Research and Technology - Hellas, IESL, Voutes, 71110 Heraklion, Greece
  \and Department of Physics and Institute for Plasma Physics, University of Crete, GR-71003 Heraklion, Greece
  \and Crimean Astrophysical Observatory, P/O Nauchny Crimea 98409, Russia
  \and Special Astrophysical Observatory of the Russian AS, Nizhnij Arkhyz, Karachaevo-Cherkesia 369167, Russia 
  \and Southern Station of the Sternberg Astronomical Institute, Moscow M.V. Lomonosov State University, P/O Nauchny, 298409, Crimea, Russia
  \and Department of Physical Sciences, Hiroshima University, Higashi-Hiroshima, Hiroshima 739-8526, Japan
  \and ZAH, Landessternwarte Heidelberg, K\"onigstuhl 12, 69117 Heidelberg, Germany
  \and Instituto de Astronom\'ia, Universidad Nacional Aut\'onoma de M\'exico, 22860 Ensenada BC, M\'exico
  \and Abastumani Observatory, Mt. Kanobili, 0301 Abastumani, Georgia
  \and Inst. of Solar-Terrestrial Physics, Lermontov st. 126a, Irkutsk p/o box 291, 664033, Russia
  \and Center for Backyard Astrophysics - New Mexico, PO Box 1351 Cloudcroft, NM 88317, USA
  \and Engelhardt Astronomical Observatory, Kazan Federal University, Tatarstan, Russia
  \and Department of Physics and Astronomy, University of Southampton, Southampton, SO17 1BJ, United Kingdom
  \and Kavli Institute for the Physics and Mathematics of the Universe (Kavli IPMU), The University of Tokyo, 5-1-5 Kashiwa-no-Ha, Kashiwa City Chiba, 277-8583, Japan
  \and Finnish Centre for Astronomy with ESO (FINCA), University of Turku, V\"ais\"al\"antie 20, FI-21500 Piikki\"o, Finland
  \and Department of Physics, University of Colorado Denver, CO, USA
  \and Steward Observatory, University of Arizona, Tucson, AZ 85721  USA
  \and Lowell Observatory, Flagstaff, AZ 86001, USA
}

\date{Received 10 November 2015; accepted 29 February 2016}

\abstract
{Over the past few years, on several occasions, large, continuous
  rotations of the electric vector position angle (EVPA) of linearly
  polarized optical emission from blazars have been reported. These
  events are often coincident with high energy $\gamma$-ray flares and
  they have attracted considerable attention, since they could allow us to
  probe the magnetic field structure in the $\gamma$-ray emitting
  region of the jet. The flat-spectrum radio quasar \object{3C 279} is one of
  the most prominent examples showing this behaviour.}
{Our goal is to study the observed EVPA rotations and to distinguish
  between a stochastic and a deterministic origin of the polarization
  variability.}
{We have combined multiple data sets of $R$-band
  photometry and optical polarimetry measurements of \object{3C 279},
  yielding exceptionally well-sampled flux density and polarization
  curves that cover a period of 2008--2012\thanks{The measured and processed 
  optical polarization and R-band photometry data are available in electronic 
  form at the CDS via anonymous ftp to cdsarc.u-strasbg.fr (130.79.128.5) or 
  via http://cdsweb.u-strasbg.fr/cgi-bin/qcat?J/A+A/.}. Several large 
  EVPA rotations are identified in the data. We introduce a quantitative
  measure for the EVPA curve smoothness, which is then used to test a set
  of simple random walk polarization variability models against the data.}
{\object{3C 279} shows different polarization variation characteristics during
  an optical low-flux state and a flaring state. The polarization
  variation during the flaring state, especially the smooth
  $\sim360\,^\circ$~rotation of the EVPA in mid-2011, is not consistent with
  the tested stochastic processes.}
 {We conclude that, during the two different optical flux states, two different
  processes govern polarization variation, which is possibly a
  stochastic process during the low-brightness state and a
  deterministic process during the flaring activity.}


\maketitle

%CONTENT%%%%%%%%%%%%%%%%%%%%%%%%%%%%%%%%%%%%%%%%%%%%%%%%%%%%%%%%%%%%%%%%%%%%%%%%

%_NEW_SECTION__________________________________________________________________
\section{Introduction}
\label{sec:intro}

Blazars, a class of active galactic nuclei containing relativistic
jets of magnetized plasma that are pointed close to our line of sight, are
known to be highly variable in the optical polarization fraction and
electric vector position angle (EVPA) since the early days of
polarimetric observations of quasars
\citep[e.g.][]{1967ApJ...148L..53K}.  Monotonic optical EVPA
rotations spanning days to months and with amplitudes of up to
several full cycles have been observed, for example by
\citet{1988A&A...190L...8K}, \citet{2008A&A...492..389L}, and
\citet{2010ApJ...710L.126M}.  However, the $n\pi$-ambiguity of EVPA
makes it difficult to detect large continuous rotations ($\geq
180\,^\circ$) in sparsely sampled polarization curves.  Therefore, a
significant monitoring effort with several participating observatories
is required to find and study these events.

During the past few years several optical EVPA rotations have been
reported as coinciding with flaring activity in $\gamma$-ray emission, as
observed by the \textit{Fermi} $\gamma$-ray Space Telescope at
GeV~energies and atmospheric imaging Cherenkov telescopes at
TeV~energies;
for instance the $\sim 240\,^\circ$~rotation in \object{BL Lac} in 2005
\citep{2008Natur.452..966M},
in \object{PKS 1510-089} the counter-clockwise $\sim 720\,^\circ$~rotation in
2009 \citep{2010ApJ...710L.126M} and the counter-clockwise $\sim 380\,^\circ$
and clockwise $\sim 250\,^\circ$~rotations in 2012 \citep{2014A&A...569A..46A},
and the rotation in \object{W Comae} \citep{2014ApJ...794...54S}.
These coincident events include rotations in the bright $\gamma$-ray quasar
\object{3C 279} \citep{2010Natur.463..919A, 2014A&A...567A..41A}.
\object{3C 279} has exhibited large EVPA rotations on both short (days) and
long (months) time-scales and with opposite senses of rotation
\citep{2008A&A...492..389L, 2010Natur.463..919A}.
The coincidence of EVPA rotations and $\gamma$-ray flares has raised
considerable interest towards optical polarization monitoring, since it may
provide clues about the origin of the high energy emission in blazars.

There are several open questions regarding the observed EVPA swings:
What physical processes produce the rotations of the EVPA?
Is it the same process for all events, in all objects, or different
processes for different events, even in the same object?
And is the $\gamma$-ray flaring activity physically connected to the
rotations of the optical polarization angle or are those events just
coincidences \citep{2015MNRAS.453.1669B}?
Various models have been proposed to explain the EVPA rotations or, more
recently, the potential connection between the EVPA rotation and
$\gamma$-ray flares.

These models can be divided into two classes:
deterministic and stochastic models.
The deterministic models include, for example, the superposition of two 
emission components with different polarization characteristics
\citep{1984MNRAS.211..497H}, the shock compression of an ordered helical
magnetic field in an axisymmetric, straight jet
\citep{2014ApJ...789...66Z,2015ApJ...804...58Z}, and several models that are
based on non-axisymmetric structures in the jet.
The latter group of models includes, for example, a global bend in the jet, 
where a change of the viewing angle $\Delta\theta$ can produce a rotation of 
the EVPA $\Delta\theta < \chi < 180\,^\circ$ owing to relativistic aberration
\citep[e.g.][]{1982ApJ...260..855B, 2010IJMPD..19..701N}.
\citet{2010Natur.463..919A} explained the 2009 swing in \object{3C 279} by a
bent jet and \citet{2014A&A...567A..41A} used this same model for the swing
in 2011.
Non-axisymmetric magnetic field configuration was invoked to explain EVPA
swings in BL Lacertae objects by \citet{1985ApJ...289..188K}, while
\citet{1988A&A...190L...8K} introduced a qualitative model of a shock passing
through a helical magnetic field with a changing pitch angle.
Recently, \citet{2008Natur.452..966M} proposed a model containing an emission
feature, which does not fill the entire jet cross-section, on a
helical trajectory that probes different parts of a large-scale helical
magnetic field as it moves along the jet.
This model has also been applied to the 2006/2007 EVPA swing in \object{3C 279}
\citep{2008A&A...492..389L}.

The second class of models is based on a stochastic variation of polarization 
parameters of multiple cells \citep[e.g.][]{1985ApJ...290..627J, 
1988ApJ...332..678J, 2007ApJ...659L.107D, 2014ApJ...780...87M}.
These studies have shown that it is easy to produce random EVPA changes that
appear as a rotation of several hundred degrees.
The two most recent models by
\citet{2014ApJ...789...66Z,2015ApJ...804...58Z} -- in the class of
deterministic models -- and \citet{2014ApJ...780...87M} -- in the class of
stochastic models -- in particular explore the connection between EVPA rotations
and $\gamma$-ray flares.

One key argument in distinguishing between deterministic and stochastic
polarization variation is the smoothness of a continuous EVPA rotation
\citep[e.g.][]{2008Natur.452..966M}.
Deterministic models should produce smooth EVPA variation whereas stochastic
models are expected to produce more erratic EVPA curves.
Here we develop a method, based on this kind of a quantitative measure of a
curve smoothness, to test the stochastic models of polarization variability and
apply it to the optical polarization data set of \object{3C 279} that was 
collected during our intensive multi-wavelength campaign in 2010--2012.

In \cref{sec:data}, we describe our almost three years of polarimetry data. We
introduce the quantitative measure of the EVPA curve smoothness and
analyse the polarization variation in \cref{sec:dataanalysis}.  In
\cref{sec:randomwalks} and \cref{sec:simresults} we test three simple
random walk processes against the observations, and these results are
discussed in \cref{sec:discussion}. A summary and conclusions are
presented in \cref{sec:conclusions}.
This publication is based on the PhD~thesis ``Origin of the $\gamma$-ray 
emission in AGN jets - A multi-wavelength photometry and polarimetry data 
analysis of the quasar \object{3C 279}'' \citep{PhDThesis}.

%_NEW_SECTION__________________________________________________________________
\section{Data}
\label{sec:data}

An intensive VLBI and multi-waveband monitoring campaign targeted on
quasar \object{3C 279}, including observatories covering radio,
millimetre, infrared, optical, ultraviolet, X-ray, and $\gamma$-ray
bands, was carried out in 2010--2012 (\textsc{Quasar Movie Project};
Kiehlmann et al., in prep.).  For the period covered by the campaign we have
accumulated $R$-band photometry and optical ($R$ and $V$ filters,
spectropolarimetry at 5000--7000\,\AA, and white light) polarimetry
data from a number of existing blazar monitoring programs and
from an {\it ad hoc} campaign specifically targeted on \object{3C 279}.
Furthermore, we include in our analysis some data taken before the
campaign period by the long-term monitoring programs, so that the data
presented in this paper cover a time range of
$\mathrm{JD}\,2454790$--$\mathrm{JD}\,2456120$.  The
observations were performed by (1) the 70\,cm telescope of Abastumani
Astrophysical Observatory (Mount Kanobili, Georgia), (2) the 2.2\,m
telescope of Calar Alto Observatory (Almer\'ia, Spain), (3) the 70\,cm
AZT-8 telescope of the Crimean Astrophysical Observatory (CrAO;
Nauchnij, Russia), (4) the 1.5\,m KANATA telescope of the
Higashi-Hiroshima Observatory (Hiroshima, Japan), (5) the 35~and 60\,cm
Kungliga Vetenskapakademien (KVA) telescopes and (6) the 2\,m Liverpool
telescope of the Observatorio del Roque de Los Muchachos (La Palma,
Canary Islands, Spain), (7) the 1.83\,m Perkins telescope of Lowell
Observatory (Flagstaff, Arizona, USA), (8) the 1\,m telescope of the
Special Astrophysical Observatory of the Russian Academy of Sciences
(SAO~RAS; Nizhny Arkhyz, Russia), (9) the 1.3\,m SMARTS telescope of
the Cerro Tololo Inter-American Observatory (Chile), (10) the 84\,cm
telescope of the Observatorio Astron\'omico Nacional (San Pedro
M\'artir, Mexico), (11) the 40\,cm LX-200 telescope of St.~Petersburg
State University (St.~Petersburg, Russia), and (12) the 1.54\,m Kuiper
and 2.3\,m Bok telescopes of Steward Observatory (Mt.~Bigelow and Kitt
Peak, Arizona, USA).  In addition, data were also gathered in the
observing campaign of the American Association of Variable Star
Observers (AAVSO), and by individual observers
(T.~Krajci\footnote{Observing as member of the Center for Backyard
  Astrophysics.} using a 36\,cm telescope and A.~Sadun using the
robotic 50\,cm New Mexico Skies (NMS) telescope~11).  The
participating observatories are listed in \cref{tab:obs}. 

\Cref{tab:obs} also includes, for most of the participating telescopes,
references to the descriptions of the programs and the used data reduction
procedures. In the following we shortly comment on the reduction of
data from the rest of the telescopes. Observations with the 1\,m
telescope of SAO~RAS, the 70\,cm AZT-8 telescope of CrAO, the 50\,cm
telescope of NMS, and by T.~Krajci were made through colour filters
intended to implement the spectral response curve close to the one of
\citet{1976MNSSA..35...70C} $R_C$ system. Various software
implementations of the differential aperture photometry technique were
applied to measure the brightness of \object{3C 279} on CCD images. Based on
the seeing conditions an appropriate circular aperture size was chosen
by each observer. The observers were advised to use the standardized
list of comparison stars maintained by LSW~Heidelberg\footnote{\url{
 http://www.lsw.uni-heidelberg.de/projects/extragalactic/charts/1253-055}},
although this was not always possible. Some
observers used ensemble photometry while others employed a single
comparison star.  In particular, the 1\,m SAO and the 70\,cm AZT-8
CrAO telescope images were reduced \citep[see][]{2005Ap.....48..156D}
using the star N9 from \citet{2001AJ....122.2055G}.

We combined all $R$-band photometry into one light curve after
cross-calibrating all data sets with respect to the SMARTS $R$-band data to
compensate for systematic offsets between different instrument and telescope
combinations and we converted magnitudes into spectral flux densities
(Kiehlmann et al., in prep.).
We removed one data point from the KVA polarimetry data, identified as an
outlier ($>60\,^ \circ$ intraday jump in EVPA) based on the combined optical
EVPA curve.
Four data points in San Pedro M\'artir (SPM) polarimetry data were identified
as outliers based on the simultaneous EVPA jumps in both \object{3C 279} and
\object{3C 273} which was observed together with \object{3C 279} in the
framework of the Quasar Movie Project.
Additionally, we removed five data points from the St.~Petersburg, four points
from the CrAO, and one point from the Liverpool telescope datasets.
These points show large uncertainties and offsets from the combined optical
polarization fraction curve, that are not typical of \object{3C 279} on
comparable time-scales.
We corrected the polarization fraction curves for the statistical (Rician) bias
following \citet{2014MNRAS.442.1693P}.
In optical the polarization fraction and the EVPA depend only weakly on the
frequency (see panels c and d in \cref{fig:3C279polopt}).
Therefore, we combine all optical polarization curves.
We follow the convention that the EVPA is measured counter-clockwise from
North.
An increase of the polarization angle refers to a counter-clockwise rotation 
projected on the sky.

Before analysing the data we averaged data points within 0.5 days in the
$R$-band light curve and in the combined optical polarization curves.
The averaging intervals are selected iteratively instead of using fixed time
bins, taking into account the uneven time sampling of the data. 
The polarization data is converted to Stokes parameters before averaging and
converted back afterwards.
The Stokes parameters are averaged weighted with the uncertainties.
The variability on time-scales smaller than half a day is of the order of the
measurement uncertainties of individual data points.
The averaging thus reduces the noise without removing significant real
variations.
The $R$-band light curve and the combined optical polarization curves are shown
in \cref{fig:3C279polopt}, panel~(b)--(d).

\begin{table}
  \caption{Observatories contributing optical polarization data to the Quasar
           Movie Project with telescope diameters, filters, number of
           polarization data points, and the type of data with references to
           the data reduction in the table footnotes.}
  \label{tab:obs}
  \centering
  \begin{tabular*}{\columnwidth}{@{\extracolsep{\fill} } L{2.4 cm} r l R{0.5 cm} L{0.4 cm} L{0.4 cm}}
    \toprule
    Obs.  &
    Tel. diam.  &
    Filter  &
    $N_\mathrm{pol}$  &
    \multicolumn{2}{l}{Data/Ref.}  \\
    \midrule
    AAVSO                       & various       & R                          &     &                        & phot\tablefootmark{4}   \\
    Abastumani,                 & 0.7\,m        & R                          &     &                        & phot\tablefootmark{5}   \\
    \ \ Georgia \\
    Calar Alto,                 & 2.2\,m        & R                          & 15   & pol\tablefootmark{6}  & phot\tablefootmark{6}   \\
    \ \ Spain\tablefootmark{1}  \\
    CrAO, Russia                & 0.7\,m        & R                          & 84   & pol\tablefootmark{7}  & phot\tablefootmark{8}   \\
    KANATA, Japan               & 1.5\,m        & V                          & 72   & pol\tablefootmark{9}  &                         \\
    T. Krajci, USA              & 0.36\,m       & R                          &      &                       & phot\tablefootmark{8}   \\
    KVA, La Palma               & 0.6\,m        & WL\tablefootmark{2}        & 14   & pol\tablefootmark{10}  &                         \\
                                & 0.35\,m       & R                          &      &                       & phot\tablefootmark{10}   \\
    Liverpool,                  & 2\,m          & V+R                        & 41   & pol\tablefootmark{10}  &                         \\
    \ \ La Palma                & 2\,m          & R                          &      &                       & phot\tablefootmark{10}   \\
    NMS T11, USA                & 0.5\,m        & R                          &      &                       & phot\tablefootmark{8}   \\
    Perkins, USA                & 1.83\,m       & R                          & 68   & pol\tablefootmark{7}  & phot\tablefootmark{7}   \\
    SAO RAS,                    & 1.0\,m        & R                          &      &                       & phot\tablefootmark{8}   \\
    \ \ Russia \\
    SMARTS, Chile               & 1.3\,m        & R                          &      &                       & phot\tablefootmark{11}   \\
    St. Petersburg,             & 0.4\,m        & R                          &      &                       & phot\tablefootmark{7}   \\
    \ \ Russia                  & 0.4\,m        & WL\tablefootmark{2}        & 49   & pol\tablefootmark{7}  &                         \\
    SPM, Mexico                 & 0.84\,m       & R                          & 46   & pol\tablefootmark{12} & phot\tablefootmark{12}  \\
    Steward Obs.,               & 1.54, 2.3\,m  & Spec.\tablefootmark{3}     & 210  & pol\tablefootmark{13} &                         \\
    \ \ USA                     & 1.54, 2.3\,m  & R                          &      &                       & phot\tablefootmark{13}  \\
    \bottomrule
  \end{tabular*}
  \tablefoot{
    \tablefoottext{1}{Calar Alto data was acquired as part of the MAPCAT project: \url{http://www.iaa.es/~iagudo/research/MAPCAT}.}
    \tablefoottext{2}{White light.}
    \tablefoottext{3}{5000--7000\,\AA, including nine data points observed in V-band.}
    \tablefoottext{4}{\url{http://www.aavso.org/}.}
    \tablefoottext{5}{\cite{Kurtanidze1}, \cite{Kurtanidze2}.}
    \tablefoottext{6}{\cite{2012IJMPS...8..299A}.}
    \tablefoottext{7}{\cite{2010ApJ...715..362J}.}
    \tablefoottext{8}{See text.}
    \tablefoottext{9}{\cite{2008SPIE.7014E..4LK}.}
    \tablefoottext{10}{\cite{2014A&A...567A..41A}.}
    \tablefoottext{11}{\cite{2012ApJ...756...13B}.}
    \tablefoottext{12}{\cite{2013ApJS..206...11S}.}
    \tablefoottext{13}{\cite{2009arXiv0912.3621S}.}
  }
\end{table}

%_NEW_SECTION__________________________________________________________________
\section{Polarization analysis}
\label{sec:dataanalysis}

Before the description and analysis of the optical light curve and the 
polarization curves we discuss a solution to the $n\pi$-ambiguity of the 
polarization angle and introduce a measure to quantify the smoothness of a 
curve.

%_NEW_SUBSECTION_______________________________________________________________
\subsection{The $n\pi$-ambiguity of the polarization vector}
\label{sec:evpashifting}

The EVPA, $\chi$, is defined within a
$180\,^\circ$-interval\footnote{The choice of the EVPA interval-limits is
arbitrary.
Usual limits are $0\,^\circ {\leq} \chi {<} 180\,^\circ$ or
$-90\,^\circ {\leq} \chi {<} 90\,^\circ$.  We use the former.}
Thus, the identification of EVPA swings is ambiguous because of the
$\pi$-modulus: $\chi = \chi \pm n\pi$, $n \in \mathbb{N}$.
This \emph{$n\pi$-ambiguity} causes problems in analysing EVPA variability,
since the difference $\Delta\chi = \chi_2 -\chi_1 \pm n\pi$ is ambiguous and
even the direction of the rotation is ambiguous
\citep[e.g.][]{2008Natur.452..966M,2008A&A...492..389L,2010Natur.463..919A}.
To visualize rotations larger than $\pi$, data points are usually shifted by
$n\pi$ minimizing the difference between adjacent data points

\begin{align}
  \chi_{i,\mathrm{adj}} = \chi_i - n \pi
  \text{\ \ \ with\ \ \ }
  n = \mathrm{int}\left( \frac{\chi_i - \chi_{i-1}}{\pi} \right),
  \label{eq:adjustEVPA1}
\end{align}

where $\mathrm{int}(\cdot)$ denotes rounding to the nearest integer.
This procedure is based on the assumption of minimal variation between adjacent
data points and relies on adequate sampling.

It has been suggested to consider the EVPA uncertainties in this procedure
\citep[e.g.][]{2011PASJ...63..489S,2013ApJS..206...11S}.
In that approach an EVPA data point $\chi_i$ is shifted according to
\cref{eq:adjustEVPA1} only if it shows a significant offset from the previous
data point $\chi_{i-1}$, i.e.
$\Delta \chi_\mathrm{red} = \left| \chi_i -\chi_{i-1} \right|
-\sqrt{e^2_{\chi_i} + e^2_{\chi_{i-1}}} > \frac{\pi}{2}$,
where $e_{\chi_i}$, $e_{\chi_{i-1}}$ are the corresponding uncertainties.
This method results in EVPA curves that may depend on the choice of the initial
two quadrants it is measured in.
We consider, for example, two data points $\chi_1=40\,^\circ$ and
$\chi_2=140\,^\circ$ in the initial interval $\left[0, 180\,^\circ\right)$,
both with uncertainty $e_\chi=10\,^\circ$.
Then, $\Delta \chi_\mathrm{red} = 85.9\,^\circ$ and $\chi_2$ is not shifted,
yielding a counter-clockwise rotation of $\chi_2 - \chi_1 = 100\,^\circ$.
Considering the same data in the initial interval
$\left[-90, 90\,^\circ\right)$, i.e. $\chi_1=40\,^\circ$ and
$\chi_2=-40\,^\circ$, gives $\Delta \chi_\mathrm{red} = 65.9\,^\circ$.
Again, $\chi_2$ is not shifted but this time yields a clockwise rotation of
$-80\,^\circ$.
This method produces results that depend on the choice of the initial
interval.
Instead, using \cref{eq:adjustEVPA1} yields a consistent result
independently of the choice of the initial interval.

In \cref{app:EVPAadjustment} we introduce a generalization of
\cref{eq:adjustEVPA1}, which considers more than a single preceding
data point, and the \emph{consistency level} as a quality check of the
adjusted EVPA curve.
Under the assumption that the adjusted EVPA curve accurately represents the
intrinsic variation of the EVPA curve, the consistency level allows one to
estimate the probability that the curve is correctly reconstructed.
A higher consistency level indicates better sampling and a more reliable
adjusted EVPA curve.

%_NEW_SUBSECTION_______________________________________________________________
\subsection{A quantitative measure of smoothness}
\label{sec:variationestimator}

To gauge the smoothness of an EVPA curve, we define a \emph{variation
estimator}, a quantitative measure of the (erratic) variability of a curve.
The larger the variation estimator, the less smooth the curve is.
First, we define the pointwise, local derivative of the EVPA curve in units of
degrees per unit time:

\begin{align}
  \left( \frac{\Delta \chi}{\Delta t} \right)_i = \frac{\chi_i - 
\chi_{i-1}}{t_i - t_{i-1}}.
  \label{eq:ptpvar}
\end{align}

The mean of the derivative
$\bar{\chi}_t = \left< \Delta \chi / \Delta t \right>$  
indicates a secular trend of the data.
The deviation of the local derivative from the secular trend is calculated at
each data point as:

\begin{align}
  s_i = \left( \frac{\Delta \chi}{\Delta t} \right)_i - \left< \left( 
\frac{\Delta \chi}{\Delta t} \right) \right>.
\end{align}

A local derivative of the order of the secular trend indicates a smooth
variation with $s_i \sim 0$. 
Deviation of the derivative from the trend, $\pm s_i > 0$, indicates erratic
variation, i.e. the curve is more jagged. 
We define the variation estimator as the mean absolute $s_i$:

\begin{align}
  s = \left< \left| s_i \right| \right> = \left< \left| \left( \frac{\Delta 
\chi}{\Delta t} \right)_i - \left< \left( \frac{\Delta \chi}{\Delta t} \right) 
\right> \right| \right>.
  \label{eq:variationestimator}
\end{align}

We use $s$ as an estimator for the smoothness of a curve with respect to a
potential linear trend.
An EVPA curve with $s_1$ is considered \emph{smoother} than a second curve
with $s_2$ if $s_2 > s_1$.

The standard deviation of the pointwise derivative
$\sqrt{\left< s_i^2 \right>}$ could be used as an alternative approach to
quantify the EVPA variability.
Averaging over $s_i^2$ increases the weight of larger deviations from the
secular trend compared to averaging over $\left| s_i \right|$, which makes the
standard deviation of $s_i$ more susceptible to outliers than the variation
estimator defined in \cref{eq:variationestimator}.
The measured variation estimator is biased by measurement errors and curvature 
of the smooth EVPA variation.
Measurement errors do not average out because of the absolute mean in
\cref{eq:variationestimator}.
Thus errors increase the variation estimator.
Only first order trends -- estimated through
$\left< \left(\Delta \chi / \Delta t \right) \right>$ -- are considered to
contribute to real EVPA rotations in this analysis.
This is the simplest assumption about the intrinsic, smooth variability.
Higher order variation needs {\it a priori} knowledge and modelling of the
variation, which is not given.
This higher order variation (curvature) increases the variation estimator,
i.e. for non-linear deterministic variation the smoothness of the curve can be
underestimated.
Both biases are discussed in more detail in \cref{app:varestbias}.

Before analysing the smoothness of the EVPA curve using the variation
estimator $s$, it is crucial to average over data points that occur in short 
time intervals, in which the measurement errors are expected to be larger
than the real variation.  EVPA offsets of the order of the errors at
small time intervals will lead to large local derivatives and
artificially increase the variation estimator~$s$.

%_NEW_SUBSECTION_______________________________________________________________
\subsection{Optical polarization variation in \object{3C 279}}
\label{sec:3C279polarization}

\begin{figure*}[p!]
\centering
\includegraphics[width=\textwidth]{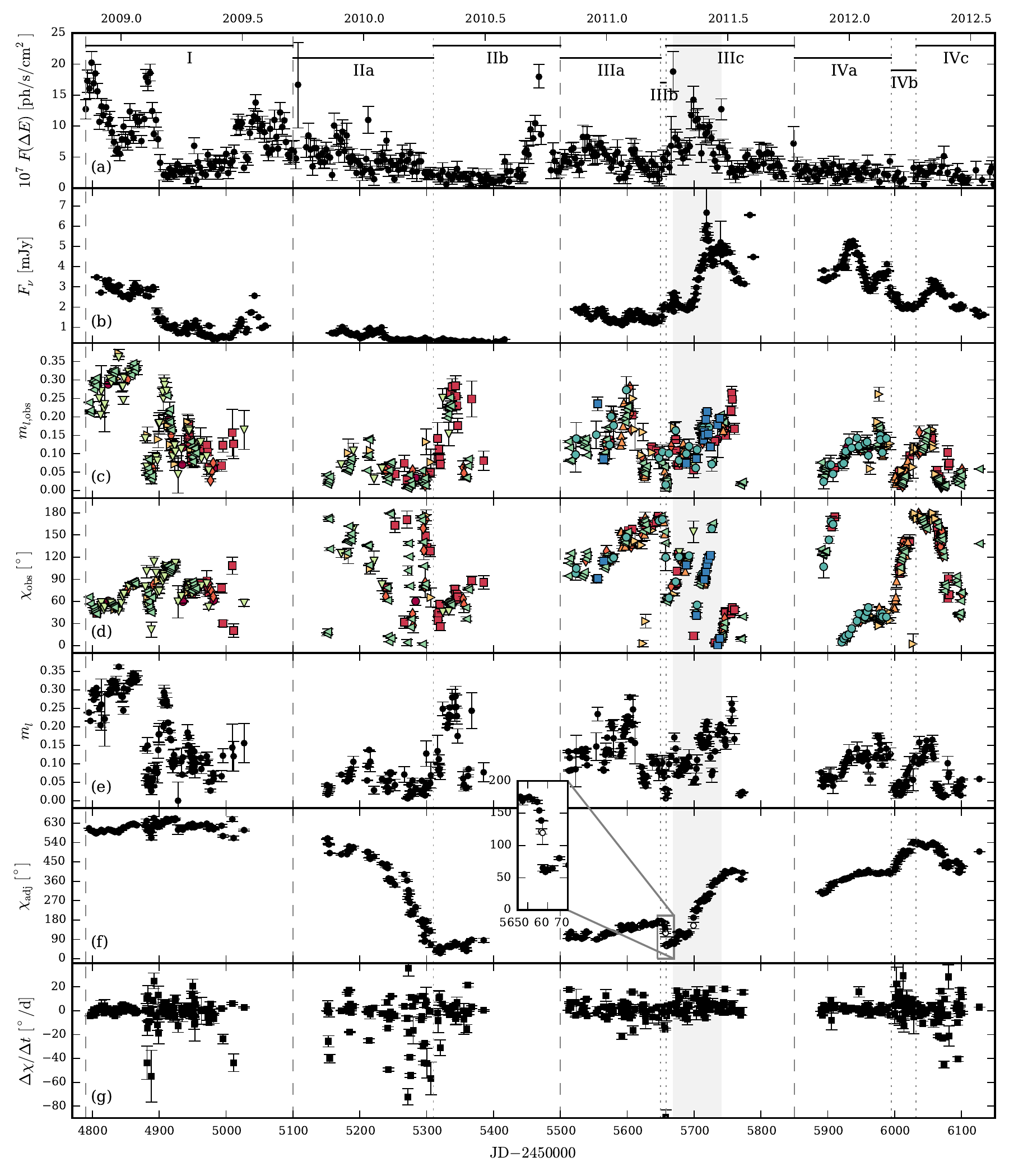}
\caption{Optical photometry and polarimetry and $\gamma$-ray light curve of
  \object{3C 279}. Fermi-LAT $\gamma$-ray light curve at $>100\,\mathrm{MeV}$
  binned into 3~day intervals (panel~a) as published in
  \citet{2015ApJ...807...79H}.
  Combined $R$-band light curve (panel~b).
  Measured, optical polarization fraction (panel~c) and EVPA (panel~d);
  red circles: Calar Alto (R),
  red squares: CrAO-70cm (R),
  red diamonds: Perkins (R),
  orange up-sided triangles: SPM (R),
  orange right-sided triangles: St.~Petersburg (R),
  green down-sided triangles: KANATA (V),
  green left-sided triangles: Steward Obs. (spec. and V),
  blue circles: Liverpool (V+R),
  blue squares: KVA (white light).
  Combined, de-biased, and averaged polarization fraction (panel~e). Combined,
  averaged, and adjusted EVPA (panel~f); open symbols are added from the
  non-averaged EVPA curve. Pointwise, local derivative of the adjusted
  EVPA (panel~g). The grey area highlights the period of $\gamma$-ray
  flaring activity coinciding with a rotation of the optical polarization
  angle.}
  \label{fig:3C279polopt}
\end{figure*}

\Cref{fig:3C279polopt} shows the combined $R$-band light curve (panel~b),
the observed optical polarization fraction (panel~c) and EVPA curves
(panel~d) from individual telescopes, the combined, debiased, and
averaged polarization fraction (panel~e), the combined, $n\pi$-adjusted,
and averaged EVPA (panel~f), and the pointwise, local derivative of the
adjusted EVPA (panel~g) of \object{3C 279} between $\mathrm{JD}\,2454790$
and $\mathrm{JD}\,2456120$.
The notable yearly gaps around August--November are caused by the Sun proximity
and split the data into four observation periods (I, II, III, IV).

During the first observing period (period~I; Oct 2008--Nov 2009) the $R$-band
light curve exhibits diminishing flaring activity.
During period~II (Nov 2009--Aug 2010) the optical light curve only has a mean 
flux density of $0.5\,\mathrm{mJy}$ with a standard deviation of
$\sigma(f_\nu) = 0.2\,\mathrm{mJy}$.
Periods~III (Nov 2010--Aug 2011) and~VI (Nov 2011--Aug 2012) show flaring
activity with the flux density ranging from
$1.1\,\mathrm{mJy}$ to $6.6\,\mathrm{mJy}$ with a mean of $2.7\,\mathrm{mJy}$.
These outbursts will be discussed in more detail in another paper
(Kiehlmann et al., in prep.).

The optical polarization fraction and the EVPA exhibit strong variability.
The polarization fraction ranges from $\sim 0$ to $0.37$ with a mean of
$\sim0.12$ and a standard deviation of $0.08$.
The EVPA shows swings in both directions with amplitudes of up to
$500\,^\circ$.
The EVPA variation looks smoother during the periods~I, III and~IV than during
period~II, as it is already evident in the $180\,^\circ$-interval, while the
variation during period~II does not show any distinct structure.
Furthermore, the lower scatter in the plot of the EVPA derivative (panel~g) 
during the periods~I, III and~IV indicates a smoother variation.

In \Cref{fig:3C279polopt}, we also mark times at which the EVPA curve shows an
abrupt change in its behaviour.
For each period of consistent EVPA change (periods~IIa to~IVc, interrupted by
two observation breaks), we measure the EVPA rotation \emph{amplitude}, given
by the difference between the minimum and maximum EVPA,
$\left| \Delta\chi \right| = \max \chi(t_i) - \min \chi(t_i)$,
the corresponding \emph{duration} of the largest rotation, i.e. the time
passing between the minimum and maximum of the EVPA curve, $\Delta t$,
the mean and standard deviation of the polarization fraction, the variation
estimator, $s$, and the consistency level of $n\pi$-adjustments,
$N_\mathrm{cons}$ (see \cref{app:EVPAadjustment}).
Errors on the polarization fraction mean and standard deviation are estimated
using a bootstrap method with 10\,000 iterations.
Furthermore, we estimate the error bias of the EVPA variation estimator for
each observing period through simulations and calculate the debiased
variation estimator for each period using \cref{eq:varestdebias}.
The results are listed in \cref{tab:polarization}.

The mean polarization fraction is low during period~IIa as compared to the
flaring periods, during which it can be up to $\sim 3$ times higher.
The standard deviation of the polarization fraction ranges from $0.028$ to
$0.084$.
There is no clear connection between the mean polarization fraction and the
standard deviation.

During period~I the EVPA shows little variation within an interval of 
93~degrees.
The time of the apparent $208\,^\circ$~rotation in 2009, reported by
\citet{2010Natur.463..919A}, is included in period~I.
We discuss the absence of this apparent rotation in our data in more detail in
\cref{sec:apparent2009rotation}.
Period~II exhibits erratic EVPA variability with an overall clockwise trend up
to an amplitude of at least $494\,^\circ$.
The low consistency level $N_\mathrm{cons}=1$ indicates that period~IIa is not
sampled sufficiently to reliably reconstruct the intrinsic EVPA variation.
The estimation of the probability of correct reconstruction ranges from $0\,\%$
to $<76\,\%$ (c.f.~\cref{app:EVPAadjustment}).

There are six periods of smooth-looking EVPA rotations, periods~IIIa--c,
and IVa--c.
The first smooth EVPA rotation during the period IIIa is an increase of
$86^\circ$ with a slow rate of $+0.9^\circ/\mathrm{d}$, which is followed by a
sharp decrease of the EVPA by $110^\circ$ with a rate of
$-16.0^\circ/\mathrm{d}$.
This distinct change in the EVPA at $\mathrm{JD}\,2455650$ takes place at the
beginning of an optical flaring period.
During the optical flaring activity in the period~IIIc, the EVPA quite smoothly
increases by $352^\circ$ over 98 days, corresponding to a mean rate of
$+3.6^\circ/\mathrm{d}$.
In the period~IVa, the EVPA rotates $109^\circ$ with a rate of
$+1.6^\circ/\mathrm{d}$ and this rotation continues in the period~IVb for
another $131^\circ$ at a faster rate of $+4.7^\circ/\mathrm{d}$.
Thereafter the sense of rotation changes again and the EVPA decreases by
$140^\circ$ at a rate of $+3.6^\circ/\mathrm{d}$.
Periods~IIIb and IIIc have low consistency levels, each owing to a sampling 
gap.
\Cref{fig:3C279polopt} shows the unaveraged data points in these gaps
as open symbols.
These data increase the consistency level, indicating a higher reliability of
the adjusted EVPA curve.
The EVPA swings during period~IVa and~IVb are well sampled
$N_\mathrm{cons}{>}50$, $N_\mathrm{cons}=23$, corresponding to correct
reconstruction probabilities of ${>}99.9\,\%$ and $63$--$99\,\%$.

The EVPA variation estimator reaches its highest values during period~IIIb.
We note that this result is strongly affected by the sparse sampling of the
rapid rotation with only nine data points.
The secular trend of the EVPA is not reliably estimated and a single data point
diverging from this trend by $\sim 80\,^\circ/\mathrm{d}$ drastically increases
the variation estimator which would otherwise be of the order of
$s_\mathrm{debiased} \sim 5\,^\circ/\mathrm{d}$.

During period~IIa, the EVPA variation estimator is significantly larger than in
the subsequent periods, confirming the observation that the EVPA variation is
more erratic during period~IIa and smoother afterwards.
The erratic variations during period~IIa can be either intrinsic or, if the
(heteroscedastic) measurement uncertainties are underestimated, owing to
measurement noise.
The latter explanation would require the measurement uncertainties during
period~IIa to be underestimated by a factor of 3.7, which is not plausible.
Therefore, we conclude that the erratic EVPA behaviour during period~IIa is
source-intrinsic.
The debiased variation estimator is $\sim 5\,^\circ/\mathrm{d}$ during
periods~IIb--IIIc, during period~IVa it is of the order of the error bias, and
it increases towards the end of the data (period~IVc) , when the flaring
activity is declining.

\begin{table*}
  \caption{Optical polarization characteristics of \object{3C 279} for
    different periods (col.~1): period time range (col.~2), mean and
    standard deviation of the polarization fraction (col.~3, 4),
    rotation amplitude over rotation duration (col.~5), total (col.~6)
    and debiased (col.~7) EVPA variation estimator, and the variation
    estimator bias (col.~8), EVPA adjustment consistency level based
    on the averaged data (col.~9) and the non-averaged data (col.~10,
    if differing).}
  \label{tab:polarization}
  \centering
  \begin{tabular*}{\textwidth}{@{\extracolsep{\fill} } r r r r r r r r r r }
    \toprule
    Period  &
    $\mathrm{JD}{-}2450000$  &
    $\left< m_l \right>$  &
    $\sigma(m_l)$  &
    $\Delta\chi/\Delta t$  &
    $s$  &
    $s_\mathrm{debiased}$  &
    $s_\mathrm{bias}$  &
    \multicolumn{2}{c}{$N_\mathrm{cons}$}  \\
    &  &  &  &
    $[^\circ/\mathrm{d}]$  &
    $[^\circ/\mathrm{d}]$  &
    $[^\circ/\mathrm{d}]$  &
    $[^\circ/\mathrm{d}]$ 
    &  &  \\
    \midrule
    tot   &             & $0.123\pm0.003$  & $0.083\pm0.002$  &             &  7.2  & $ 6.4\pm0.4$  & $ 3.2\pm0.4$  & 1         & (2)   \\
    I     & 4790--5100  & $0.184\pm0.007$  & $0.100\pm0.003$  &             &  6.1  & $ 5.0\pm1.3$  & $ 3.6\pm1.3$  & 101       & (138) \\
    II    & 5100--5500  & $0.089\pm0.007$  & $0.077\pm0.005$  &             & 13.9  & $13.4\pm1.2$  & $ 3.6\pm1.2$  & 1         & (2)   \\
    III   & 5500--5800  & $0.132\pm0.004$  & $0.058\pm0.002$  &             &  5.7  & $ 5.1\pm0.5$  & $ 2.5\pm0.5$  & 2         & (6)   \\
    IV    & 5800--6120  & $0.080\pm0.003$  & $0.046\pm0.001$  &             &  5.5  & $ 4.2\pm0.9$  & $ 3.5\pm0.9$  & 22        & (24)  \\
    \midrule
    IIa   & 5100--5310  & $0.050\pm0.003$  & $0.030\pm0.003$  & $-494/154$  & 17.8  & $17.4\pm1.2$  & $ 3.6\pm1.2$  & 1         & (2)   \\
    IIb   & 5310--5500  & $0.164\pm0.013$  & $0.084\pm0.004$  & $  62/ 47$  &  6.2  & $ 5.0\pm1.2$  & $ 3.6\pm1.2$  & $\infty$  &       \\
    IIIa  & 5500--5650  & $0.139\pm0.006$  & $0.056\pm0.003$  & $  86/ 93$  &  4.8  & $ 4.0\pm0.5$  & $ 2.5\pm0.5$  & $\infty$  & (77)  \\
    IIIb  & 5650--5658  & $0.060\pm0.011$  & $0.029\pm0.008$  & $-110/  7$  & 26.0  & $25.9\pm0.5$  & $ 2.5\pm0.5$  & 3         & (8)   \\
    IIIc  & 5658--5850  & $0.131\pm0.006$  & $0.057\pm0.003$  & $ 352/ 98$  &  4.8  & $ 4.0\pm0.5$  & $ 2.5\pm0.5$  & 2         & (6)   \\
    IVa   & 5850--5995  & $0.104\pm0.004$  & $0.034\pm0.002$  & $ 109/ 69$  &  2.1  & $ \approx 0$  & $ 3.5\pm0.9$  & 50        & (64)  \\
    IVb   & 5995--6032  & $0.045\pm0.005$  & $0.028\pm0.004$  & $ 131/ 28$  &  7.1  & $ 6.2\pm0.9$  & $ 3.5\pm0.9$  & 23        & (30)  \\
    IVc   & 6032--6120  & $0.073\pm0.006$  & $0.050\pm0.002$  & $-140/ 39$  &  8.0  & $ 7.2\pm0.9$  & $ 3.5\pm0.9$  & 22        & (24)  \\
    \bottomrule
  \end{tabular*}
  \tablefoot{A consistency level of $\infty$ implies that all data points lie within a $90\,^\circ$-interval.}
\end{table*}

%_NEW_SECTION__________________________________________________________________
\subsection{The apparent $208\,^\circ$~rotation in 2009}
\label{sec:apparent2009rotation}

\begin{figure}[h]
  \centering
  \includegraphics[width=\columnwidth]{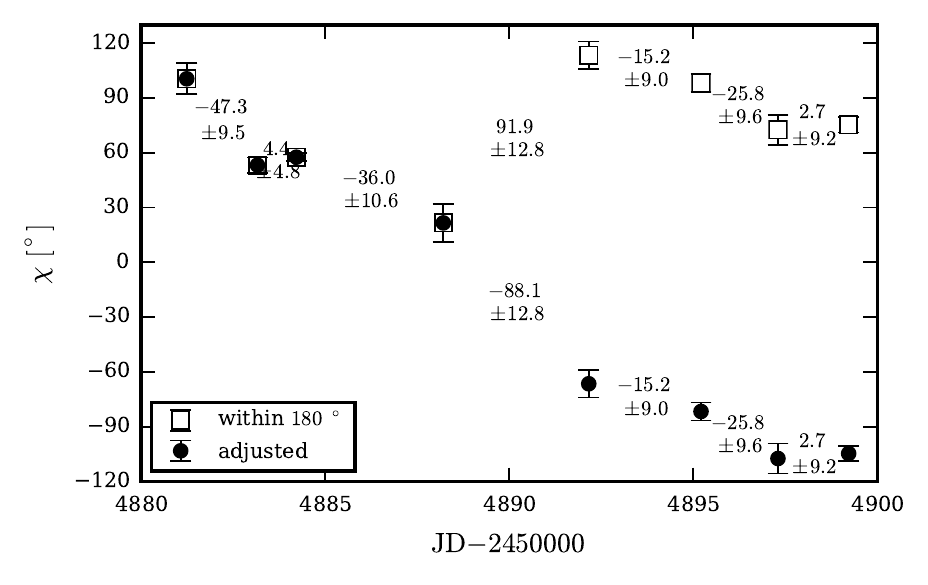}
  \caption{The $208\,^\circ$ EVPA rotation presented in
           \citet{2010Natur.463..919A} based on data from the KANATA telescope.
           Open squares show the EVPAs within the initial
           $0$--$180\,^\circ$-interval, black circles the adjusted EVPA curve.
           The numbers show the difference between adjacent data points and the
           corresponding uncertainties in degrees.}
  \label{fig:kanata}
\end{figure}

\citet{2010Natur.463..919A} report an apparent $208\,^\circ$~clockwise rotation
of the EVPA in \object{3C 279} from $\mathrm{JD}\,2454880$ to
$\mathrm{JD}\,2454900$ based on KANATA data shown in \cref{fig:kanata}.
The larger than $180\,^\circ$~rotation is a result of the usual scheme of
minimizing differences between adjacent data points
(c.f.~\cref{eq:adjustEVPA1}).
The rotation is sparsely sampled, particularly, there is one gap close to
$90\,^\circ$ between $\mathrm{JD}\,2454888$ and $\mathrm{JD}\,2454892$.
Within the uncertainty of $\sigma_{\Delta\chi} = 12.8\,^\circ$ it is unclear
whether this difference corresponds to a clockwise or counter-clockwise
rotation.

\begin{figure}[ht]
  \centering
  \includegraphics[width=\columnwidth]{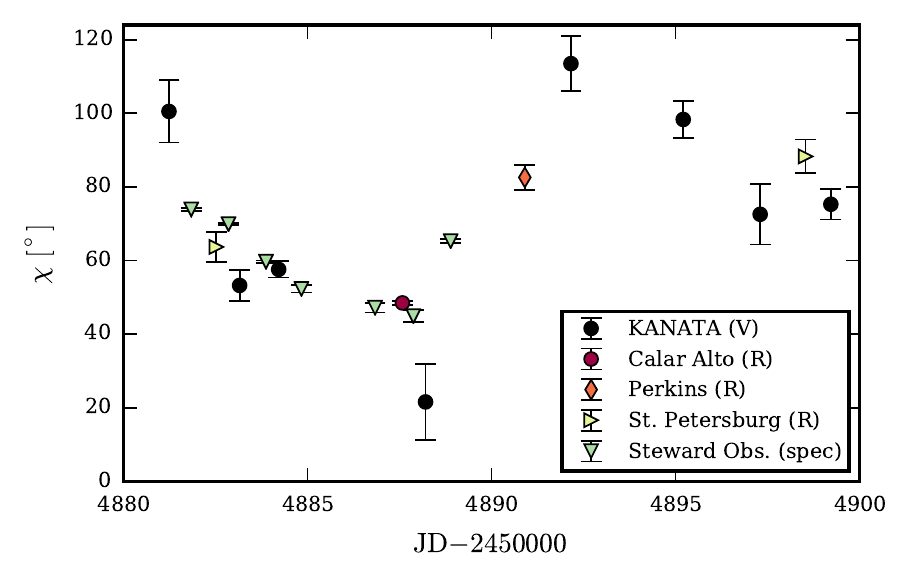}
  \caption{The EVPA data as shown in \cref{fig:kanata} with additional
           data from Calar Alto, Perkins, St. Petersburg, and Steward Obs.}
  \label{fig:qmp}
\end{figure}

\Cref{fig:qmp} shows additional data from four more telescopes.
Additional data from Calar Alto, Perkins, St. Petersburg and Steward Obs.
indicate changes of the rotation direction and variability within less than
$\sim90\,^\circ$, instead of a continuous, large, clockwise rotation.
The KANATA data point at $\mathrm{JD}\,2454888$ appears to be a potential
outlier.
Omitting this data point, the EVPA only varies within $\sim 60\,^\circ$.

The reported continuous, clockwise rotation of $208\,^\circ$ cannot
be maintained.
This example clearly demonstrates that densely sampled data is necessary to
treat the $n\pi$-ambiguity.

%_NEW_SECTION__________________________________________________________________
\subsection{Optical polarization variability and $\gamma$-ray activity}
\label{sec:gamma}

For a comparison with the optical light curves, panel~a in
\cref{fig:3C279polopt} shows the Fermi-LAT $\gamma$-ray light curve at
$0.1$--$300\,\mathrm{GeV}$ binned into 3~day intervals.
This light curve was originally published in \citet{2015ApJ...807...79H}.
\object{3C 279} exhibits strong $\gamma$-ray flaring activity 
($>10^{-6}\,\mathrm{ph\,s^{-1}\,cm^{-2}}$) at various occasions.
During the first period of high $\gamma$-ray activity around
$\mathrm{JD}\,2454800$ the optical EVPA is roughly constant.
During the prominent flare around $\mathrm{JD}\,2454890$ reported by
\citet{2010Natur.463..919A}, we have shown that the EVPA varies within less
than $90\,^\circ$, though the polarization fraction drops significantly.
Two high activity states coincide with observation gaps in the optical
data.
The high $\gamma$-ray activity state at $\sim \mathrm{JD}\,2455666$--$2455741$,
showing multiple peaks, coincides exactly with the $352\,^\circ$~rotation of
the optical polarization angle during period~IIIc and flaring activity at
optical bands.
Although the EVPA rotation and optical flaring activity continues during
periods~IVa to IVc, $\gamma$-ray activity is low
($<4 \cdot 10^{-7}\,\mathrm{ph\,s^{-1}\,cm^{-2}}$).
In contrast to the $352\,^\circ$~rotation of the optical EVPA coinciding with
strong $\gamma$-ray activity, the large rotation during period~IIa only 
coincides with mild $\gamma$-ray activity
($\lesssim 10^{-6}\,\mathrm{ph\,s^{-1}\,cm^{-2}}$).
We do not find an obvious, consistent correspondence between the optical
polarization variability and $\gamma$-ray flaring activity.
However, the fact that the $352\,^\circ$~EVPA rotation during the period~IIIc
exactly brackets a long-term $\gamma$-ray active period, strongly suggests that
this particular polarization angle rotation event is connected to the
$\gamma$-ray emission.

%_NEW_SECTION__________________________________________________________________
\section{Random walk models}
\label{sec:randomwalks}

The prevailing physical scenario for launching relativistic jets
\citep[e.g.][]{2007MNRAS.380...51K,2011MNRAS.418L..79T} involves a strong,
outwardly propagating, helical magnetic field.
The flow of plasma in the jet accelerates and becomes increasingly collimated
with distance from the black hole, with the magnetic energy density decreasing
in favour of a higher kinetic energy density.
It is likely that current-driven instabilities disrupt the helical ordering
near end of the jet's acceleration zone \citep[e.g.][]{2012MNRAS.427.2480N}.
This, plus transverse velocity gradients in the flow
\citep[e.g.][]{2004ApJ...605..656V} probably result in the flow becoming
turbulent.
Electrons can be energized by second-order Fermi acceleration
\citep{2008ApJ...681.1725S} and magnetic reconnections in the turbulent plasma
\citep{2015MNRAS.450..183S}.
We model this turbulence in terms of cells, each of which contains a uniform
magnetic field, although the actual geometry is certainly more complex than
this.
We choose the field direction randomly from one cell to the next.
Although more sophisticated schemes have been used involving nested cells of
different sizes \citep{1988ApJ...332..678J,2015IAUS..313..122M}, here we limit
our analysis to the basic case of independent cells.

We investigate whether the observed polarization variation characteristics
($\left<m_l\right>$, $\sigma(m_l)$, $\Delta\chi$, and $s$) can be produced by
stochastic processes.
We do so by performing random walks in Stokes parameters $I$, $Q$, and $U$, and
comparing the properties of the obtained polarization curves to the observed
ones.
The model consists of $N_\mathrm{cells}$ cells.
The properties of a sub-set of cells are changed at each time step.
The sampling of the total simulation time $T$ is randomized with time
steps $\Delta t$ following a truncated power law distribution
$P(\Delta t) \propto \Delta t^\alpha$ with $\alpha {<} -1$ within limits
$\left[\Delta t_\mathrm{min}, \Delta t_\mathrm{max}\right]$.
The parameters best describing the distribution of time steps in each
observation period are shown in \cref{tab:secmodelpars}, col.~2--3.
The distribution limits are directly adopted from the data.
We derive $\alpha_t$ in the following manner:
(i) from the observation time steps we calculate the maximum-likelihood
estimator (MLE) of the power law index $\hat{\alpha}_{t,\mathrm{obs}}$,
(ii) we simulate time steps testing different values of $\alpha_t$ and
calculate the MLE power law index of the simulated time steps 
$\hat{\alpha}_{t,\mathrm{sim}}$, (iii) we use the value of $\alpha_t$ for which
$\hat{\alpha}_{t,\mathrm{obs}}=\hat{\alpha}_{t,\mathrm{sim}}$.
The variation rate $n_\mathrm{var}$ sets the number of cells that change their
properties per unit time step (one day).
The total number of cells changed between time $t_i$ and $t_{i-1}$ is given by
$N_\mathrm{var}(t_i, t_{i-1}) = \mathrm{int}\left(n_\mathrm{var} t_i\right) -
\mathrm{int}\left(n_\mathrm{var} t_{i-1}\right)$, where $\mathrm{int}(\cdot)$
denotes rounding to an integer.
If $N_\mathrm{var}(t_i, t_{i-1}) > N_\mathrm{cells}$ all cells are changed
during that time step.
In the following we define three simple random walk processes, that differ in
the properties of the cells and in the selection of cells be to changed.

%___NEW_SECTION________________________________________________________________
\subsection{Simple Q,U random walk process}
\label{sec:simpleQU}

We create $N_\mathrm{cells}$ initial cells with uniform intensity $I_i$.
Each cell has a uniform magnetic field oriented randomly.
The EVPA is oriented accordingly.
The cell size, thus, corresponds to the largest scale of uniform magnetic
field.
Uniformity implies that each cell is maximally polarized.
The polarization fraction $m_{l,\mathrm{max}}$ of synchrotron radiation in a
uniform magnetic field is \citep[p. 217]{Longair201103}

\begin{align}
  m_{l,\mathrm{max}} = \frac{p - 1}{p - \frac{7}{3}},
\end{align}

where $p$ is the power law index of the electron energy distribution.
The maximum polarization is $m_{l,\mathrm{max}} \approx 0.72$ for $p = 2.5$
\citep[p. 217]{Longair201103}.
The initial random variables $\hat{Q}_i$ and $\hat{U}_i$ are drawn from a
Gaussian distribution for each cell $i = 1 \dots N_\mathrm{cells}$.
The Stokes parameters $Q_i$, $U_i$ for each cell are obtained through the
following normalization which yields the same intensity and the maximum
polarization for each cell:

\begin{align}
  \hat{Q}_i &\sim \mathcal{N}(0, 1),
  \label{eq:randomQ}    \\
  \hat{U}_i &\sim \mathcal{N}(0, 1),
  \label{eq:randomU}    \\
  I_i &= \frac{I}{N_\mathrm{cells}} = \mathrm{const},
  \label{eq:constI}    \\
  Q_i &= \frac{\hat{Q}_i}{\sqrt{\hat{Q}_i^2 + \hat{U}_i^2}} \cdot I_i \cdot m_{l, \mathrm{max}},
  \label{eq:normQ}  \\
  U_i &= \frac{\hat{U}_i}{\sqrt{\hat{Q}_i^2 + \hat{U}_i^2}} \cdot I_i \cdot m_{l, \mathrm{max}}.
\label{eq:normU}
\end{align}

The polarization fraction and angle are independent of the total
intensity $I$ and we set $I=1$.
For the simple random walk process the $N_\mathrm{var}(t_i, t_{i-1})$ cells
that change are randomly selected every time step.
This implies that a single cell may have changed several times between
$t_{i-1}$ and $t_i$ and that the total number of different cells that change is
$\leq N_\mathrm{var}(t_i, t_{i-1})$.
The properties of the selected cells are newly set following
\crefrange{eq:randomQ}{eq:normU}.
In the following we abbreviate this simple random walk process \emph{siQU}.

\citet{1985ApJ...290..627J} note that choosing $\hat{Q}_i$ and $\hat{U}_i$ from
a uniform distribution in the interval $[-1,1]$ produces almost identical
results.
We point out that this prior selection leads to preferred directions of the
EVPA every 45~degrees and not a uniform EVPA distribution.

%___NEW_SECTION________________________________________________________________
\subsection{Ordered $Q$,$U$ random walk process with constant $I$}
\label{sec:shockQUconst}

This process implements the basic concept of a disturbance passing through a
turbulent medium, locally increasing the emissivity and successively
highlighting part of the medium.
In this case, the model cells are numbered.
The first cell represents the disturbance and all the following cells are a
trailing afterglow.
The initial cells are created exactly as in the simple $Q$,$U$ random walk
process, but the subsequent cell selection differs.
Assuming that $N(t_i)$ cells change between time $t_{i-1}$ and $t_i$, the
properties of all cells are shifted by $N(t_i)$ cells.
The first $N(t_i)$ cells, representing the region right behind the disturbance,
are newly set with random properties.
The properties of the last $N(t_i)$ trailing cells vanish.
We call this process \emph{orQUc} in the following.
The orQUc process is less randomized than the siQU process.
Each cell the disturbance passes through has its properties maintained for a
given amount of time.
For the siQU process, on the other hand, the properties of a single cell are
maintained a random amount of time depending on when it is randomly selected to
change.

%___NEW_SECTION________________________________________________________________
\subsection{Ordered $Q$,$U$ random walk process with decreasing $I$}

For this random walk process we assume the intensity in a cell is increased by
a disturbance and gradually degrades as the disturbance moves on.
The highest intensity of the emitting region is located at the disturbance and
decreases as function of distance from the disturbance.
We scale the cell intensities $I_i$ linearly between the highest intensity at
the disturbance (cell one) and zero intensity at the
$N_\mathrm{cells}{+}1^\text{th}$ trailing cell:

\begin{align}
  I_i &= \frac{2I}{N_\mathrm{cells}^2 + N_\mathrm{cells}} (N_\mathrm{cells} -i).
  \label{eq:decrI}
\end{align}

The total intensity is again set to $I{=}1$. The cell properties are
set by \cref{eq:randomQ,eq:randomU,eq:normQ,eq:normU,eq:decrI}.  Cell
properties are changed following the scheme of the orQUc process
(\cref{sec:shockQUconst}).
We denote this process \emph{orQUd}.

The three random walk processes are simplistic implementations of a finite
emission region in a turbulent flow that is either randomly changing its
magnetic field structure (siQU) or is energized by a disturbance moving
through the plasma (orQUc, orQUd).
These implementations face several limitations.
The siQU process does not include any structure information, the individual
cells do not have a specific position.
The ordered random walk processes assume the cells are all in line, whereas
multiple cells could be next to one another.
This would affect the intensity scaling of the orQUd cells.
All processes neglect that the magnetic field orientation cannot be entirely
disconnected between different cells.
Furthermore, a disturbance such as a shock would locally order the magnetic
field and leave the field orientation in trailing cells correlated to some
extent.
Time delays which would arise from the different positions of the cells are not
considered and the modelling is performed in the observers frame under the
assumption that neither the speed nor the direction of the motion of the
emitting region changes during the polarization event.
Relativistic aberration, which would make randomly oriented cells appear to be
more aligned than they intrinsically are, is also not considered here.

%___NEW_SECTION________________________________________________________________
\subsection{Integrated polarization}
\label{sec:integratedpol}

The integrated Stokes parameters $I = \sum_{i=1}^{N_\mathrm{cells}} I_i$ and
$Q$, $U$, accordingly, determine the integrated intensity $I$, the linear
polarization fraction $m_l$ and the EVPA $\chi$ at each simulation time step:

\begin{align}
  m_l &= \frac{\sqrt{Q^2 + U^2}}{I}, \\
  \chi &= \frac{1}{2} \arctan \frac{U}{Q} + n \frac{\pi}{2} 
  \text{\ \ \ with\ \ \ } n =
  \begin{cases}
      1, & \text{if}\ Q<0 \\
      0, & \text{otherwise}
    \end{cases}
\end{align}

Since the plasma is optically thin at optical frequencies and the
Faraday rotation scales with the squared wavelength of the radiation, we can
safely ignore the effects of synchrotron self-absorption and Faraday rotation
on the polarized signal at optical frequencies.

%___NEW_SECTION________________________________________________________________
\subsection{Simulated EVPA errors}
\label{sec:errors}

\begin{figure}
  \centering
  \includegraphics[width=\columnwidth]{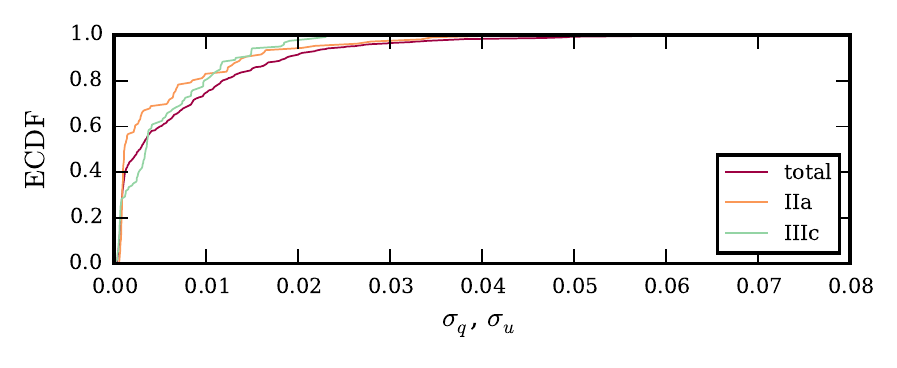}
  \caption{Empirical cumulative distribution function (ECDF) of the combined
           Stokes $q$ and $u$ uncertainties in the entire data (red curve),
           period~IIa (orange curve), and period~IIIc (green curve).}
  \label{fig:stokesuncecdf}
\end{figure}

We have tested two schemes of implementing the simulation of observational
uncertainties.
In the first scheme, we apply uncertainties to the integrated Stokes parameters
$q = Q/I$, $u = Q/I$.
For each simulated data point the uncertainties $\sigma_{q,i}$ are randomly
drawn from the empirical cumulative distribution function (ECDF) of the
combined $q$- and $u$-uncertainties measured in the corresponding period
(\cref{fig:stokesuncecdf}).
In the simulation we set $\sigma_{u,i} = \sigma_{q,i}$.
Noise $q_{\mathrm{err},i}$, $u_{\mathrm{err},i}$ is then drawn from a Gaussian
distribution $\mathcal{N}(0, \sigma_{q,i})$ and it is added to each random walk
data point.
The data points and uncertainties are then transformed into $m_l$, $\chi$ and
the corresponding uncertainties following \citet{2014MNRAS.442.1706K},
eqs.~5-6.
In the observed data the uncertainties in $q$ and $u$ is generally not the
same, but they scatter around this linear correlation.
As a consequence this method does not exactly reproduce the measured
distribution of EVPA uncertainties.

In the second scheme we apply uncertainties directly to the EVPA.
The measured EVPA uncertainties can be described by a log-normal distribution.
The distribution parameters $\mu_\mathrm{unc}$ and $\sigma_\mathrm{unc}$ are
estimated for each period individually.
To simulate EVPA ``measurement'' errors we draw a random-number
$\sigma_{\chi_i}$ from a log-normal distribution
$\mathcal{LN}(\mu_\mathrm{unc}, \sigma_\mathrm{unc})$ representing the
measurement uncertainty of each simulated data point
$\chi_{\mathrm{rw},i} = \chi_\mathrm{rw}(t_i)$.
To each data point we then add noise $\chi_{\mathrm{err},i}$ drawn from a
Gaussian distribution $\mathcal{N}(0, \sigma_{\chi_i})$ with zero mean and
standard deviation $\sigma_{\chi_i}$.
Finally, each simulated EVPA data point is given by
$\chi_{\mathrm{sim},i} = \chi_{\mathrm{rw},i} + \chi_{\mathrm{err},i}$.
This scheme produces EVPA uncertainties more close to the observed distribution
than the first scheme.
Limitation of this approach are that uncertainties in the polarization fraction
are neglected and that the simulated EVPA uncertainties are independent of the
simulated polarization fraction.

We ran all simulations presented in the following sections with both schemes to
test if the results depend on the specific implementation of the uncertainty
simulation.
All results presented are based on the first scheme, applying uncertainties to
the Stokes parameters.
Whereas the absolute numbers, distributions and figures shown in the following
slightly differ using the second scheme, the results and conclusions do not
depend on the choice of the uncertainty implementation.

%___NEW_SECTION________________________________________________________________
\subsection{Random walk simulations}
\label{sec:simulations}

We vary the \emph{primary} input parameters in the ranges
$N_\mathrm{cells} \in [2, 600]$,
$n_\mathrm{var} \in [0.1\,\mathrm{d}^{-1}, 100\,\mathrm{d}^{-1}]$.
We run the simulations in two modes.
In the first one, the cell number and the cell variation rate are randomly
drawn from (discrete/continuous) uniform distributions:
$N_\mathrm{cells} \sim \mathcal{U}_\mathrm{int}(2, 600)$ and
$n_\mathrm{var} \sim \mathcal{U}(0.1, 100.0)$.
We run a total of 1\,000\,000 simulations for each random walk process.
Additionally we run simulations with certain chosen input parameters.
For each parameter combination and random walk process we run up to 10\,000
simulations.
The choice of input parameters is described in the following sections.

The fixed \emph{secondary} input parameters for the random walk simulations are
taken from \cref{tab:secmodelpars}.
The total simulation time $T$ equals the duration of the corresponding period.
For each simulation we measure the mean polarization fraction
$\left< m_l \right>$ and its standard deviation $\sigma(m_l)$, the EVPA
amplitude $\left|\Delta\chi\right|$, and the EVPA variation estimator $s$.

\begin{table}
  \centering
  \caption{Truncated power law distribution parameters modelling the 
           random time steps (col.~2--4).}
  \begin{tabular*}{\columnwidth}{@{\extracolsep{\fill} } r | r r r }
    \toprule
    Period  &
    $\alpha_t$  &
    $\Delta t_\mathrm{min}$ $[\mathrm{d}]$  &
    $\Delta t_\mathrm{max}$ $[\mathrm{d}]$  \\
    \midrule
    total  & -1.8  & 0.5  & 21.0  \\
    II     & -1.8  & 0.7  & 21.0  \\
    III    & -1.8  & 0.5  & 14.0  \\
    \bottomrule
  \end{tabular*}
  \label{tab:secmodelpars}
\end{table}

%___NEW_SECTION________________________________________________________________
\section{Random walk simulation results}
\label{sec:simresults}

In this section we first describe general results of the random walk
simulations: expectation values of the mean and standard deviation of the
polarization fraction, a general comparison of the three random walk processes,
and dependencies between the model parameters and various measured parameters.
Then, we test the periods of the two largest rotations in the observed data
(period~IIa and IIIc) against the models.
Finally, we test the rotation amplitude distribution of the entire observed
EVPA curve against the random walk models.

%___NEW_SECTION________________________________________________________________
\subsection{Expectation values of the fractional polarization}
\label{sec:polexpectation}

For each random walk model we run simulations for selected pairs of cell
numbers and variation rates to test the expectation values of the mean and
standard deviation of the polarization fraction.
The fixed parameters are taken from \cref{tab:secmodelpars}, period~``total''.
The simulation time is $T=260\,\mathrm{d}$, corresponding to the average
observation period between sun gaps.
We run 10\,000 simulations per a pair of parameters and a random walk model.

\begin{figure}
  \centering
  \includegraphics[width=\columnwidth]{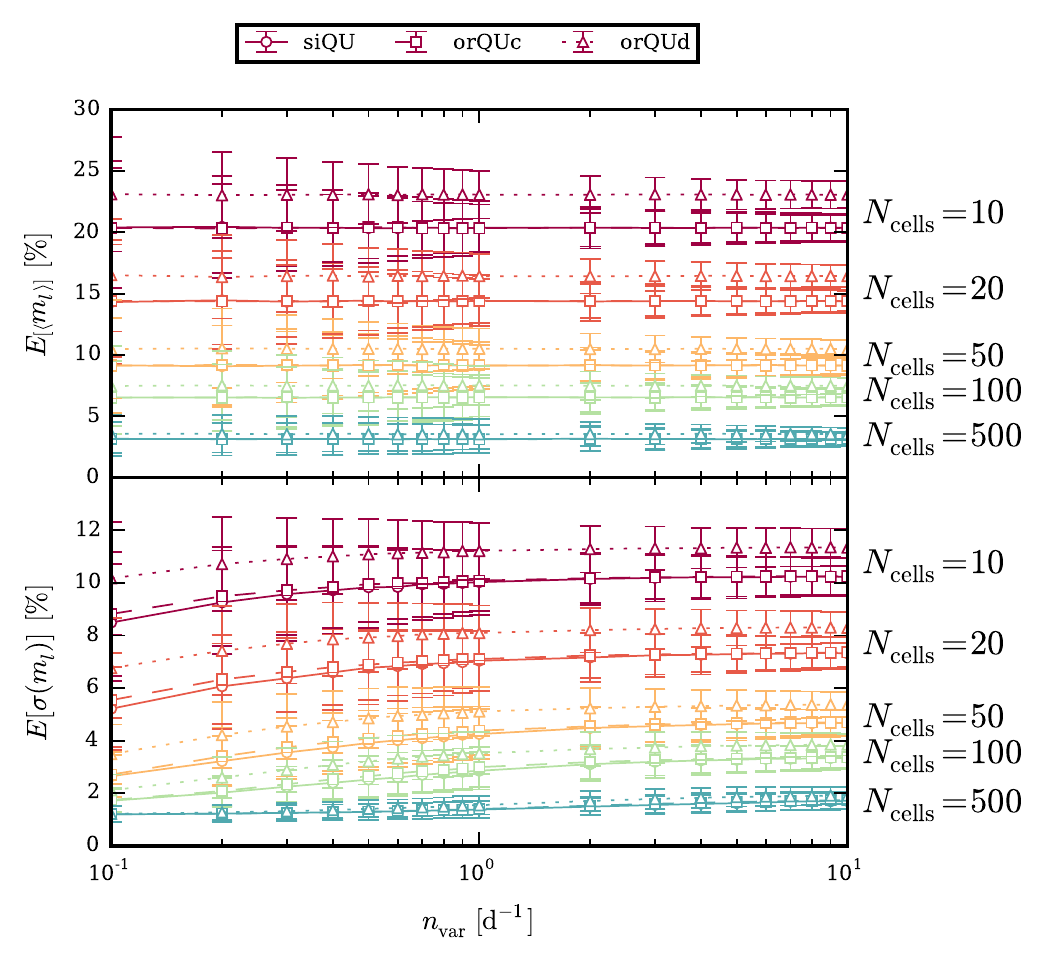}
  \caption{Expectation values of the polarization fraction mean (upper panel) 
           and standard deviation (lower panel) for different numbers of cells
           depending on the cell variation rate. Each data point is based on
           10\,000 simulations. Solid curves and circles correspond to the siQU
           model, dashed curves and squares to the orQUc model, and dotted
           curves and triangles to the orQUd model.}
  \label{fig:rwpolexp}
\end{figure}

The mean and standard deviation of the simulated polarization fraction follow
log-normal distributions for each input parameter combination.
We estimate the distribution parameters via maximum likelihood and from those
we calculate the expectation value and variance for
$\left<m_l\right>$ and $\sigma(m_l)$, for each combination of
$N_\mathrm{cells}$ and $n_\mathrm{var}$.
The results are shown in \cref{fig:rwpolexp}.
The expectation value of the mean of the polarization fraction is independent
of $n_\mathrm{var}$.

The siQU and the orQUc process produce the same amount of polarization.
Simulated curves from the orQUd process are on average more polarized by a
factor of 1.13--1.15.
As anticipated, for all processes the expected mean of the polarization
fraction depends on the cell number by
$E[\left<m_l\right>] \propto 1/\sqrt{N_\mathrm{cells}}$.
The expectation value of the standard deviation of the polarization fraction
increases with increasing $n_\mathrm{var}$ and decreases with increasing
$N_\mathrm{cells}$.
For high variation rates it saturates
at $E[\sigma(m_l)] \xrightarrow{n_\mathrm{var} \rightarrow
N_\mathrm{cells}/\Delta t} \frac{1}{2} E[\left<m_l\right>]$.

Typically, similar relations connecting the cell number and cell variation rate
with the expectation values of the polarization fraction mean and standard
deviation are used to fix the model parameters based on the observed
polarization fraction, when testing stochastic models against the data
\citep[e.g.][]{2007ApJ...659L.107D}.
We do not recommend to fix these parameters, as each pair of model parameters
results in a broad distribution of $\left<m_l\right>$ and $\sigma(m_l)$ with
standard deviations up to $\mathrm{Var}(\left<m_l\right>)^\frac{1}{2} = 0.058$
and $\mathrm{Var}(\sigma(m_l))^\frac{1}{2} = 0.024$.
The distribution is also right-tailed.
Thus, the expectation value does not match with the distribution mode.
Furthermore, the parameter combination most likely producing the observed
polarization fraction may not be the best choice for producing the observed
EVPA variability.
Instead, we recommend to test a large parameter space and to identify the
parameter region that is most likely to produce the observed polarization
fraction {\it and} the EVPA characteristics.

%___NEW_SECTION________________________________________________________________
\subsection{Model comparison and parameter dependencies}
\label{sec:modelcomparison}

\begin{figure*}
  \centering
  \includegraphics[width=\textwidth]{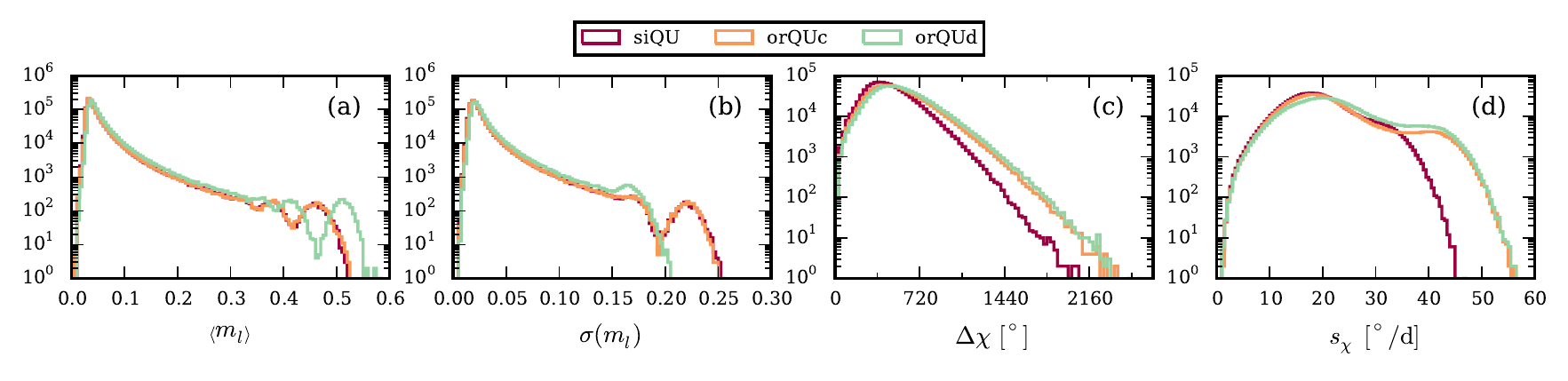}
  \caption{Distributions of (a) the mean and (b) the standard deviation of the
           polarization fraction, (c) the EVPA amplitude, and (d) the EVPA
           variation estimator for each random walk process with a range of
           model parameters.}
  \label{fig:rwcomparison}
\end{figure*}

\Cref{fig:rwcomparison} shows the distribution of the mean and standard
deviation of the polarization fraction and the EVPA amplitude and variation
estimator for each tested random walk process with primary model parameters
in the ranges $N_\mathrm{cells} \in [2, 600]$,
$n_\mathrm{var} \in [0.1\,\mathrm{d}^{-1}, 100\,\mathrm{d}^{-1}]$, secondary
parameters as listed in \cref{tab:secmodelpars}, period~``total'', and a total
time of $T=260\,\mathrm{d}$.
Each distribution is based on 1\,000\,000~simulations.
The distributions are affected, especially at the lowest values, by the limits
of the tested parameter space, and the EVPA amplitude distribution additionally
depends on the simulation total time.

The comparison of the three random walk processes based on \cref{fig:rwpolexp}
and \cref{fig:rwcomparison} shows, that the siQU process is less variable than
the orQU process, resulting on average in smaller EVPA amplitudes but smoother
EVPA curves.
The reason is, that in the siQU process single cells may change multiple times
within one time step, thus the number of cells that stay the same is larger
than in the ordered random walk processes.
The orQUd process is generally more polarized and more variable, resulting in
larger EVPA amplitudes and more erratic EVPA behaviour.
Given the intensity scaling of the cells in the orQUd process, this process is
comparable to the orQUc process with \emph{de facto} fewer cells.

In general, increasing the cell number decreases the mean and standard
deviation of the polarization fraction, the average EVPA rotation amplitude and
the variation estimator, while increasing the cell variation rate increases the
distribution means of all measured parameters.
Consequentially, there is a correlation between the EVPA rotation amplitude and
the variation estimator in the simulations.
Larger EVPA rotations are indeed on average less smooth, as one would
intuitively expect.

The results furthermore show that in principle very smooth EVPA curves
$s\sim 1\,^\circ/\mathrm{d}$ and very large rotations
$\Delta \chi \gg 360\,^\circ$ can be produced by a random walk process.
However these properties are to some extent mutually exclusive.
Therefore, it is mandatory to simultaneously consider both the rotation
amplitude and the quantified smoothness in the analysis of the random walk
simulations.

%___NEW_SECTION________________________________________________________________
\subsection{Probabilities of observed polarization variability}
\label{sec:probabilities}

The main aim of these simulations is to determine the probability of producing
the observed polarization variability characteristics by a random walk process.
We use $\left< m_l \right>$ and $\sigma(m_l)$ to characterize the polarization
fraction variation, while $\Delta\chi$ and $s$ are used to characterize the
EVPA variation.
In the following, we define four test conditions.

First, we define two polarization fraction conditions (PF1, PF2):
We search for simulations with $\left< m_l \right>$ and $\sigma(m_l)$ in
ranges $[x{-}\Delta x, x{+}\Delta x]$, where $x$ is the corresponding observed
value and $\Delta x$ is equal to the scatter of the observed
$\left< m_l \right>$ and $\sigma(m_l)$ (c.f.~\cref{tab:polarization}).
We note, that the interval range is an arbitrary choice and the final
probabilities depend on the accepted range of parameters.
Our choice is to set the accepted parameter range based on the data instead of
preselecting a fixed value.

Second, we define two polarization angle conditions (PA1, PA2):
We search for EVPA rotations with amplitudes as large or larger than observed
$\Delta\chi \geq \Delta\chi_\mathrm{obs}$ which are at the same time as smooth
or smoother than the observed data, $s \leq s_\mathrm{obs}$.
Finally, we search for simulations fulfilling all four conditions
(PF1+PF2+PA1+PA2).

We run simulations for various input parameter combinations
$N_{\mathrm{cells}, i}$, $n_{\mathrm{var}, j}$.
For each parameter combination $ij$ we run a set of $N_\mathrm{sim}{=}10\,000$
simulations.
From each simulation set $ij$ we find the number of simulations in agreement
with our conditions, $N_{ij,\mathrm{cond}}$.
The probability of producing a polarization curve in agreement with this
condition from a random walk process with input parameters
($N_{\mathrm{cells}, i}$, $n_{\mathrm{var}, j}$) is
$P_{i,j}(\mathrm{cond}) = N_{i,j,\mathrm{cond}}/N_\mathrm{sim}$.
The highest probability is $P_\mathrm{max}(\mathrm{cond}) =
\max\limits_{i,j} P_{i,j}(\mathrm{cond})$ with the optimal input parameters
($N_\mathrm{cells}^\mathrm{opt}$, $n_\mathrm{var}^\mathrm{opt}$).
The accuracy of this probability estimation depends on the sampling of the
input parameters.

Large rotations of the polarization angle are of particular interest.
Therefore we test the random walk simulations against the two largest rotations
in our data, observed during the periods~IIa and~IIIc.
In particular, we test period~IIIc because of its contemporaneity with strong
$\gamma$-ray flaring activity.
We compare these two periods as they show very different behaviour in the
smoothness of the EVPA rotation, the polarization fraction and the total flux.
% These two rotations have the highest uncertainty compared to the other periods
% in being correct representations of the intrinsic EVPA variability, as
% suggested by the low consistency level (c.f. \cref{tab:polarization}).
% This is a consequence of the large rotation amplitudes or rather the rotation
% rates, given the data sampling.
% The same uncertainty is reflected in the simulations by simulating the same
% time sampling and EVPA uncertainties as in the data.
We discuss the other periods qualitatively after the detailed study of
periods~IIa and IIIc.

%___NEW_SECTION________________________________________________________________
\subsubsection{Period IIa}
\label{sec:periodIIa}

We sample the input parameter space in the ranges
$N_\mathrm{cells}=[80, 90, \dots, 260]$ and
$n_\mathrm{var}=[4, 8, \dots, 60]\,\mathrm{d}^{-1}$.
For each parameter pair we run 10\,000 simulations.
The fixed parameters are taken from \cref{tab:secmodelpars}, period~II.
The simulation time is $T=154\,\mathrm{d}$.
The simulation selection conditions are:

\begin{alignat*}{6}
  &\text{Polarization fraction\ } &&\text{PF1:\ }  &&0.047  &&\leq  &&\makebox[\widthof{$\sigma(m_l)$}][c]{$\left<m_l\right>$}  &&\leq 0.053,                    \\
  &                               &&\text{PF2:\ }  &&0.027  &&\leq  &&\sigma(m_l)                                               &&\leq 0.033,                    \\
  &\text{Polarization angle}      &&\text{PA1:\ }  &&       &&      &&\makebox[\widthof{$\sigma(m_l)$}][c]{$\Delta\chi$}        &&\geq 494\,^\circ,              \\
  &                               &&\text{PA2:\ }  &&       &&      &&\makebox[\widthof{$\sigma(m_l)$}][c]{s}                   &&\leq 17.8\,^\circ/\mathrm{d}.
\end{alignat*}

\begin{figure}
  \centering
  \includegraphics[width=\columnwidth]{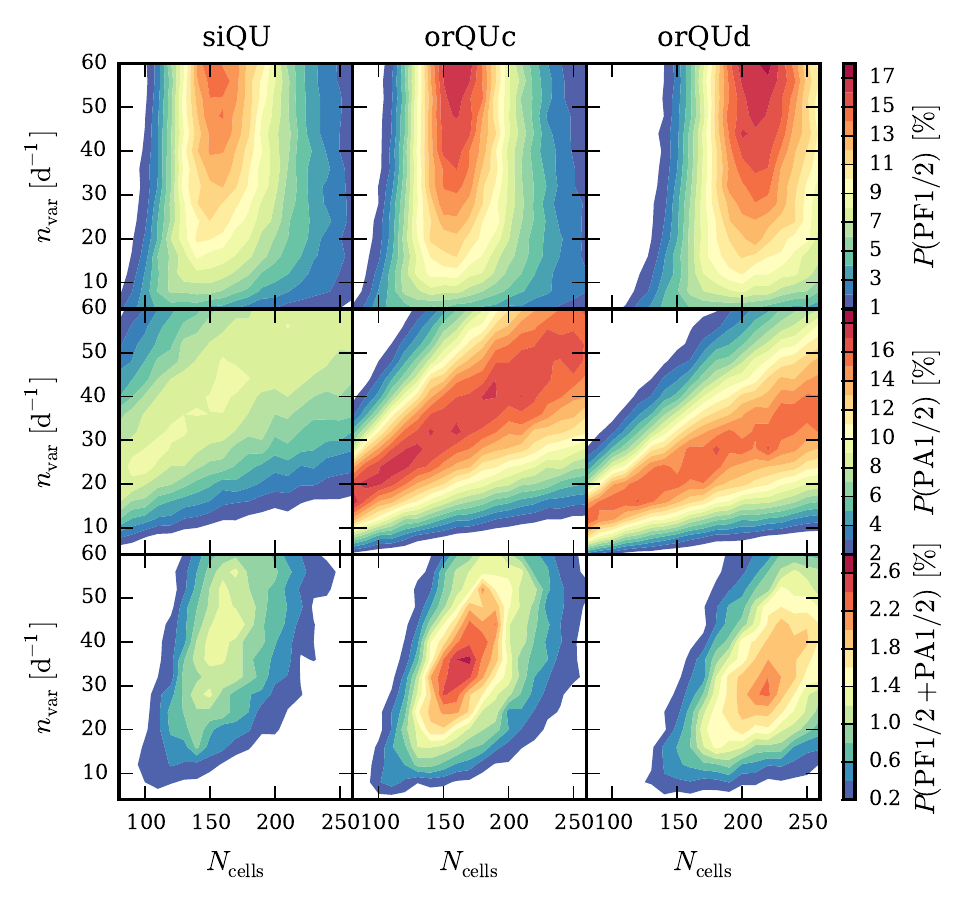}
  \caption{Probability distributions over the parameter space
           $N_\mathrm{cells}$, $n_\mathrm{var}$ for three different conditions
           (row~1--3) based on three random walk processes (column~1--3) for
           period~IIa.}
  \label{fig:p2aprob}
\end{figure}

\Cref{fig:p2aprob} shows the probability distributions over the input parameter
space for the polarization fraction conditions (PF1/2, first row), the EVPA
conditions (PA1/2, second row), and for all four conditions (PF1/2+PA1/2, third
row) for each of the three random walk processes.
The third row of \cref{fig:p2aprob} indicates that the tested parameter space
is sufficiently large to capture the region of highest probability given all
four conditions.
The highest probabilities and the corresponding optimal input parameters are
listed in \cref{tab:p2aprob}.

\begin{table*}
  \caption{Highest probabilities of producing polarization curves
           fulfilling the conditions on the polarization fraction (PF1/2), on
           the EVPA variation (PA1/2) and fulfilling all conditions from three
           different random walk processes with the corresponding, optimal
           model parameters for period~IIa.}
  \label{tab:p2aprob}
  \begin{tabular*}{\textwidth}{@{\extracolsep{\fill} } l r r r | r r r | r r r}
    \toprule
    Process  & $P(\mathrm{PF1/2})$        & $N^\mathrm{opt}_\mathrm{cells}$  & $n^\mathrm{opt}_\mathrm{var}$
             & $P(\mathrm{PA1/2})$        & $N^\mathrm{opt}_\mathrm{cells}$  & $n^\mathrm{opt}_\mathrm{var}$
             & $P(\mathrm{PF1/2+PA1/2})$  & $N^\mathrm{opt}_\mathrm{cells}$  & $n^\mathrm{opt}_\mathrm{var}$    \\
    \midrule
    siQU     & $15.4\,\%$  & $150\pm5$  & $60\pm2$  &  $9.7\,\%$  & $170\pm5$  & $44\pm2$  & $1.4\,\%$   & $170\pm5$   & $44\pm2$    \\
    orQUc    & $16.7\,\%$  & $160\pm5$  & $52\pm2$  & $17.8\,\%$  &  $80\pm5$  & $16\pm2$  & $2.7\,\%$   & $170\pm5$   & $36\pm2$    \\
    orQUd    & $17.5\,\%$  & $220\pm5$  & $60\pm2$  & $16.5\,\%$  &  $80\pm5$  & $12\pm2$  & $2.3\,\%$   & $220\pm5$   & $28\pm2$    \\
  \bottomrule
  \end{tabular*}
\end{table*}

The highest probability of producing the observed polarization fraction
characteristics (conditions PF1/2) is $\sim 18\,\%$.
It is relatively low because the observed standard deviation of the
polarization fraction of $0.03$ is larger than the maximum value of
$\sigma(m_l)$ expected from a random walk producing a mean polarization
fraction of $0.05$, which is $0.025$.
EVPA rotations of equal or higher amplitude and comparably smooth or smoother
than observed (conditions PA1/2) are occurring with probabilities up to
$10$--$18\,\%$ in the tested parameter space.
The probability of producing the observed variability in both the polarization
fraction and angle (conditions PF1/2+PA1/2) is relatively low compared to the
individual conditions.
The reason is that the comparably high standard deviation of the polarization
fraction requires higher cell variation rates than the observed EVPA
variability.
The discussed processes are capable, however, of producing the observed
polarization characteristics with a probability up to $\sim 3\,\%$.

\begin{figure}
  \centering
  \includegraphics[width=\columnwidth]{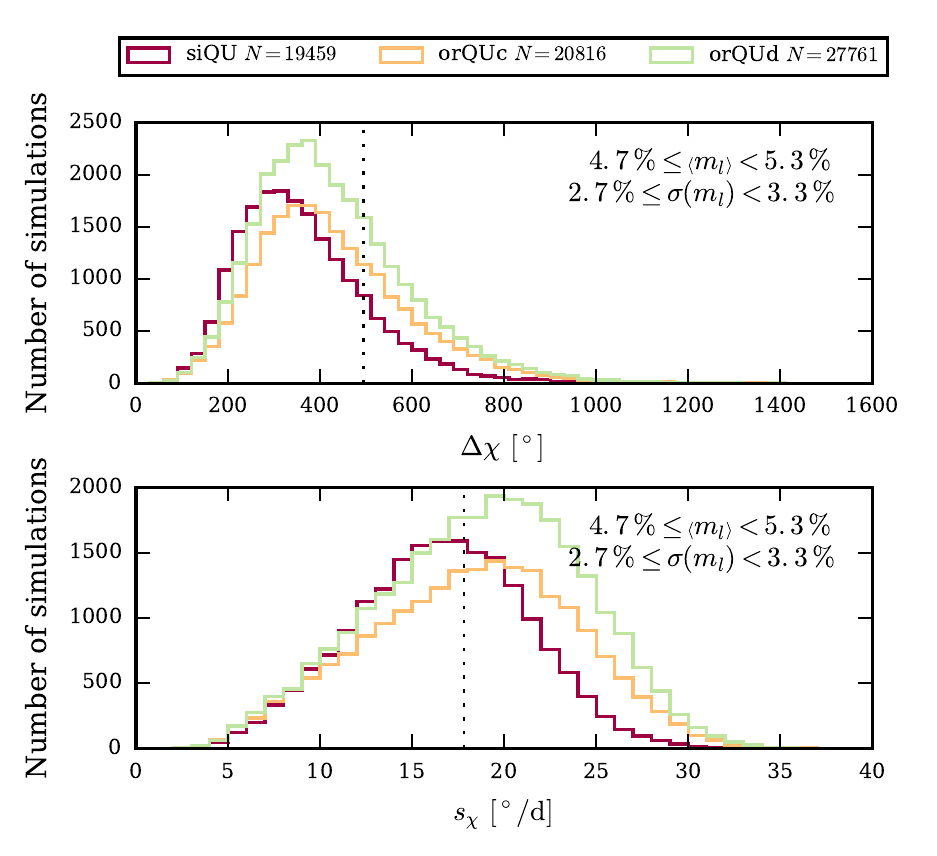}
  \caption{Distribution of the EVPA rotation amplitude (upper panel) 
           and distribution of the EVPA variation estimator (lower panel) for
           all simulations with a mean polarization fraction consistent with
           the observed value during period~IIa. The number of selected
           simulations is indicated in the legend. The corresponding observed
           values are indicated by black dotted lines.}
  \label{fig:p2adist}
\end{figure}

\Cref{fig:p2adist} shows the distribution of the EVPA rotation amplitudes
(upper panel) and the EVPA variability estimator (lower panel) for those
simulations fulfilling the polarization fraction conditions (PF1/2).
The histograms are based on 1\,000\,000 simulations per random walk process
with random primary model parameters.
The observed values during period~IIa are marked by vertical lines.
The distribution of rotation amplitudes shows that rotations as large as
observed commonly occur in random walk processes, given the observed
polarization fraction.
The observed EVPA variability estimator is close to the distribution mode.
The erratic behaviour of the EVPA during the period~IIa is characteristic of a
stochastic process.

%___NEW_SECTION________________________________________________________________
\subsubsection{Period IIIc}
\label{sec:periodIIIc}

\begin{figure}
  \centering
  \includegraphics[width=\columnwidth]{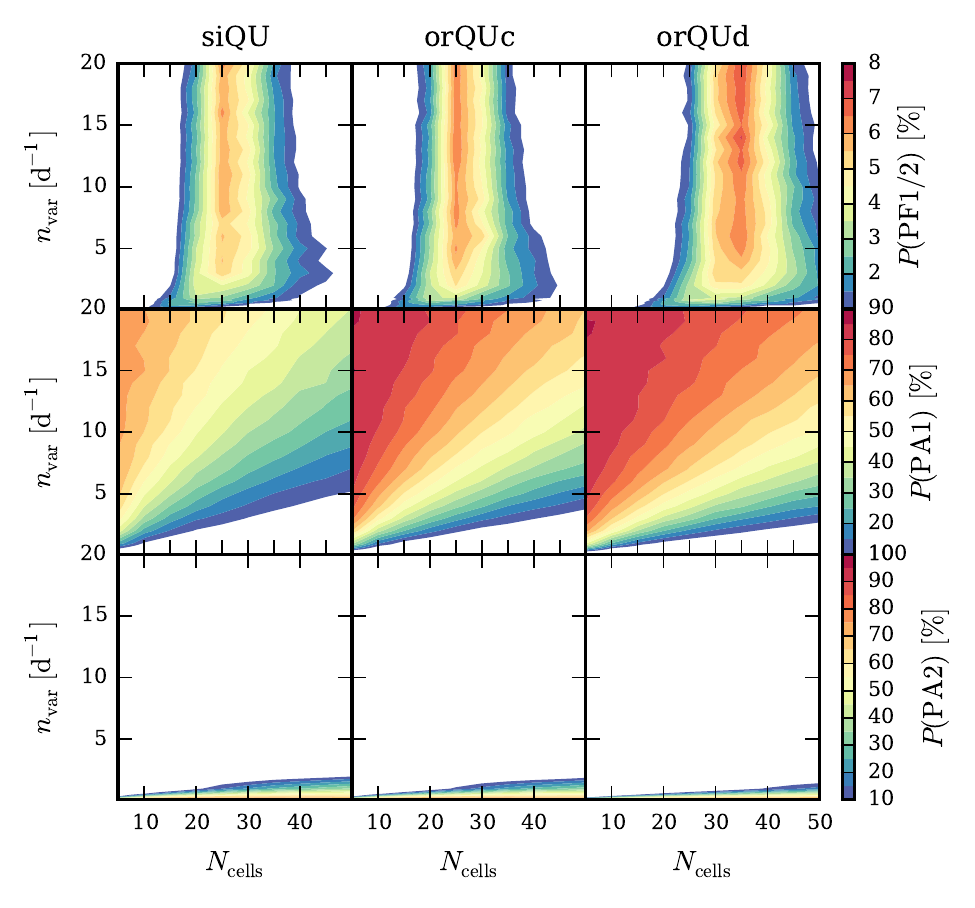}
  \caption{Probability distributions over the parameter space
           $N_\mathrm{cells}$, $n_\mathrm{var}$ for three different conditions
           (row~1--3) based on three random walk processes (column~1--3) for
           the period~IIIc.}
  \label{fig:p3cprob}
\end{figure}

To test the random walk processes against the period~IIIc data we sample the
cell number in the range $N_\mathrm{cells} = [5, 10, \dots, 50]$ and the cell
variation rate in the range
$n_\mathrm{var} = [0.1, 0.2, \dots, 1.0, 2.0, \dots, 20]\,\mathrm{d}^{-1}$.
The secondary parameters are listed in \cref{tab:secmodelpars}, period~III and
the simulation time is 98~days.
We run 10\,000 simulations per pair of parameters and use the following
conditions to analyse the simulations:

\begin{alignat*}{6}
  &\text{Polarization fraction\ } &&\text{PF1:\ }  &&0.125  &&\leq  &&\makebox[\widthof{$\sigma(m_l)$}][c]{$\left<m_l\right>$}  &&\leq 0.137,                    \\
  &                               &&\text{PF2:\ }  &&0.054  &&\leq  &&\sigma(m_l)                                               &&\leq 0.060,                    \\
  &\text{Polarization angle}      &&\text{PA1:\ }  &&       &&      &&\makebox[\widthof{$\sigma(m_l)$}][c]{$\Delta\chi$}        &&\geq 352\,^\circ,              \\
  &                               &&\text{PA2:\ }  &&       &&      &&\makebox[\widthof{$\sigma(m_l)$}][c]{s}                   &&\leq 4.8\,^\circ/\mathrm{d}.
\end{alignat*}

\Cref{fig:p3cprob} shows the probability distribution over the tested parameter
space for all three random walk processes (columns~1--3) and different
selection conditions.
The probability of producing the observed polarization fraction characteristics
(conditions PF1/2) during period~IIIc is shown in the first row and ranges up
to $\sim 7\,\%$ (c.f.~\cref{tab:p3cprob}).
The observed standard deviation of the polarization fraction of $0.057$ is
relatively low compared to the value $\sim 0.065$ expected from a random walk
with a mean polarization of $0.131$.
Thus, the probability of producing the observed mean and standard deviation is
relatively low.

The second row of \cref{fig:p3cprob} shows the probability of producing EVPA
rotations with amplitudes at least as large as the observed rotation
(condition PA1).
Those rotations are most likely produced by a few cells and high cell variation
rates.
On the other hand, the observed low EVPA variation estimator, which indicates a
smooth EVPA curve, only can be produced by random walk models with low cell
variation rates and preferentially many cells as shown in the third row of
\cref{fig:p3cprob} (condition PA2).
Consequentially, the probability of producing an EVPA rotation at least as
large and at least as smooth (conditions PA1/2) is low
(c.f.~\cref{tab:p3cprob}).

\begin{table*}
  \caption{Highest probabilities of producing polarization curves
           fulfilling the conditions on the polarization fraction (PF1/2), on
           the EVPA variation (PA1/2) and fulfilling all conditions from three
           different random walk processes with the corresponding, optimal
           model parameters for the period~IIIc.}
  \label{tab:p3cprob}
  \begin{tabular*}{\textwidth}{@{\extracolsep{\fill} } l r r r | r r r | r}
    \toprule
    Process  & $P(\mathrm{PF1/2})$        & $N^\mathrm{opt}_\mathrm{cells}$  & $n^\mathrm{opt}_\mathrm{var}$
             & $P(\mathrm{PA1/2})$        & $N^\mathrm{opt}_\mathrm{cells}$  & $n^\mathrm{opt}_\mathrm{var}$
             & PF1/2+PA1/2  \\
    \midrule
    siQU     & $6.3\,\%$  & $25\pm3$  & $20\pm0.5$  & $0.03\,\%$  & $5\pm3$  & $0.2\pm0.05$  & 0 occurrences in 10\,000 simulations  \\
    orQUc    & $6.5\,\%$  & $25\pm3$  & $13\pm0.5$  & $0.11\,\%$  & $5\pm3$  & $0.3\pm0.05$  & 0 occurrences in 10\,000 simulations  \\
    orQUd    & $7.1\,\%$  & $35\pm3$  & $14\pm0.5$  & $0.03\,\%$  & $5\pm3$  & $0.2\pm0.05$  & 0 occurrences in 10\,000 simulations  \\
  \bottomrule
  \end{tabular*}
\end{table*}

\begin{figure}
  \centering
  \includegraphics[width=\columnwidth]{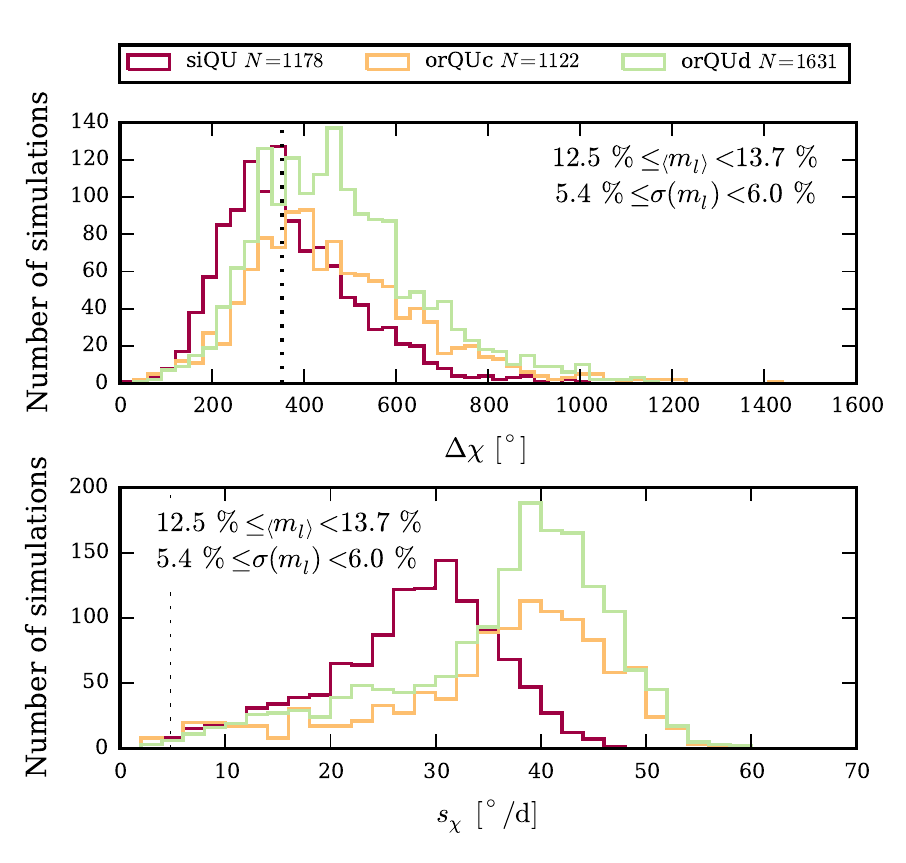}
  \caption{Distribution of the EVPA rotation amplitude (upper panel) 
           and distribution of the EVPA variation estimator (lower panel) for
           all simulations with a mean polarization fraction consistent with
           the observed value during the period~IIIc. The number of selected
           simulations is indicated in the legend. The corresponding observed
           values are indicated by black dotted lines.}
  \label{fig:p3cdist}
\end{figure}

In \cref{fig:p3cdist} we show the distribution of the rotation amplitude (upper
panel) and the EVPA variation estimator (lower panel) based on the simulations
which show the observed polarization fraction characteristics
(conditions PF1/2).
On one hand, the observed rotation amplitude is close to the distribution mode. 
On the other hand, the EVPA variability coming from a stochastic process would
be expected to be far more erratic.
Comparably smooth or smoother variations than observed do occur, but only
rarely and, most importantly, those simulations do not produce large rotation
amplitudes.
Consequentially, we do not find a single occurrence of variability comparable
to the observed period~IIIc in both the polarization fraction and angle
(conditions PF1/2+PA1/2) at any tested point in the parameter grid.
Therefore, we conclude that it is very unlikely that a stochastic process could
produce the observed EVPA rotation during the period~IIIc.

%___NEW_SECTION________________________________________________________________
\subsubsection{Other periods}
\label{sec:periodsrest}

We have focused on the two most prominent rotations during the periods~IIa and
IIIc when testing the stochastic models.
In this section we discuss the other periods qualitatively.
During period~I the EVPA shows little variability ($\lesssim 90\,^\circ$).
Though not explicitly tested for the polarization fraction of period~I, the
tests of periods~IIa and IIIc suggest that such small EVPA amplitudes are
unlikely to occur from the tested random walk processes
(c.f. \cref{fig:p2adist,fig:p3cdist}), given that the variability in the
polarization fraction is the highest during period~I.
Consistent with the likely non-stochastic period~IIIc, the polarization
fraction and the total flux is mostly high compared to period~IIa.
The average polarization fraction drops and the EVPA variability
becomes more erratic with the fading flare.
If both periods are indeed of deterministic origin, the smooth and weak EVPA
variability during period~I on the one hand, and the smooth and strong
variability during period~IIIc on the other hand, imply a complex behaviour of
\object{3C 279}.

Period~IIb shows a low-brightness state in $R$-band. Nevertheless, the
polarization fraction is exceptionally high in the mean and standard
deviation and the EVPA variation estimator is similarly low as during
the flaring states. Whether this drastic increase in the polarization
fraction marks the beginning of the following flaring state, without
showing up yet in the total flux density, cannot be properly tested
because of the following observation gap.

The polarization fraction and the EVPA variation estimator during
period~IIIa are comparable to period~IIIc despite the significantly
lower rotation rate.  \Cref{fig:p3cdist} suggests that we would expect
a larger rotation amplitude from a random walk model, arguing against
a stochastic process.  Claiming a non-stochastic origin for
periods~IIIa and IIIc implies that also period~IIIb should be produced
by another process rather than a random walk.  If all periods IIIa-c
originate in the same event, the sharp clockwise rotation opposite
to the counter-clockwise direction of the enclosing rotations poses a
challenge to any non-stochastic model.
Although not explicitly tested we suspect period~IVa not to be consistent with
the random walk processes, given the continuous, smooth rotation.

%___NEW_SECTION________________________________________________________________
\subsection{EVPA rotation amplitude distribution}
\label{sec:amplitudedistribution}

\begin{figure}
  \centering
  \includegraphics[width=\columnwidth]{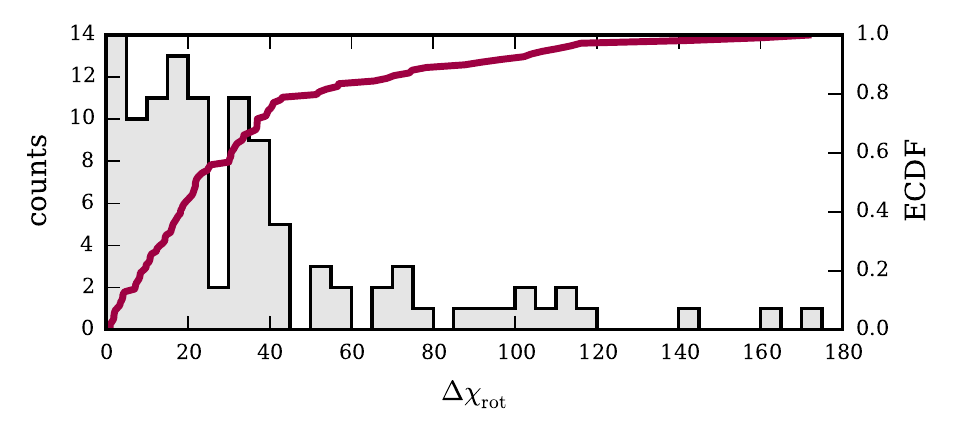}
  \caption{Histogram of the rotation amplitudes of the 109 rotations identified
           in the optical EVPA curve at $\varsigma{=}3$-significance and
           the corresponding cumulative distribution function (ECDF, red
           curve).}
  \label{fig:rotdistdata}
\end{figure}

In \cref{app:rotident} we introduce a method to automatically identify
continuous rotations in an EVPA curve based on the strict criterion that a 
continuous rotation begins and ends with a significant change of the rotation
direction at a certain significance level $\varsigma$.
We point out that the rotations discussed thus far do not strictly follow this
criterion, but were instead defined as the maximum change of the EVPA within a
certain time period that shows a general trend in the EVPA.
However, in this section we follow the strict definition.
At a $\varsigma{=}3$ significance level 109~rotations are identified in the
optical EVPA curve.
The distribution of the rotation amplitudes is shown in \cref{fig:rotdistdata}.
In \cref{app:rotident:distribution} we discuss characteristics of the rotation
amplitude distribution originating from the random walk models.

To test the observed distribution of rotation amplitudes against the random
walk models we run long simulations with a total time $T=100\,000\,\mathrm{d}$
varying the model parameters in the ranges
$N_{\mathrm{cells},i} \in [10, 20, \dots, 300]$, $n_{\mathrm{var},j} \in
[2, 4, \dots, 20]\,\mathrm{d}^{-1}$ for each random walk process.
Secondary model parameters are taken from \cref{tab:secmodelpars},
period~``total''.
For each simulation the rotations are identified at $\varsigma{=}3$
significance level and the distribution of the rotation amplitudes is tested
against the rotation amplitude distribution of the entire EVPA curve using a
two-sample Kolmogorov–Smirnov (KS) test.
We note, that this analysis is not feasible for the individual periods as too
few identified rotations do not allow us to estimate the distribution of
rotation amplitudes reliably.

\begin{figure}
  \centering
  \includegraphics[width=\columnwidth]{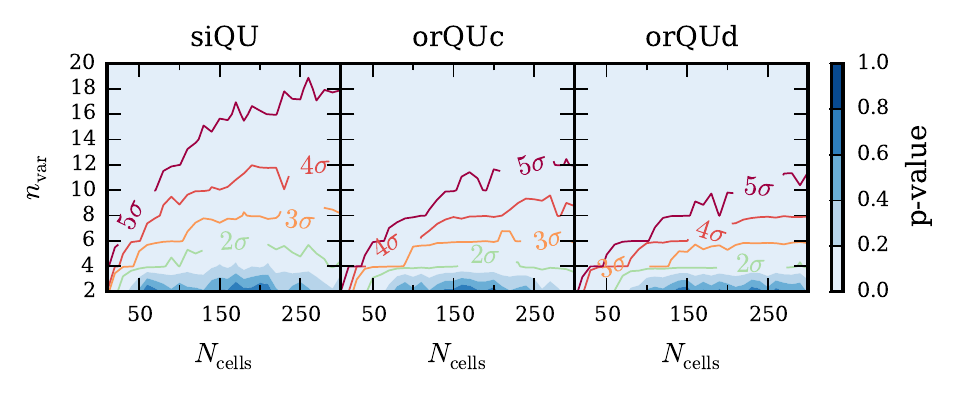}
  \caption{P-values of the KS test over a grid of model parameters, testing the
           rotation amplitude distribution of the entire EVPA curve against
           random walk simulations for three different stochastic processes.}
  \label{fig:rotdistkstest}
\end{figure}

The p-values of the KS tests are shown in \cref{fig:rotdistkstest} over the
tested parameter space for each random walk process.
Solid curves mark the edge of the parameter space where the hypothesis that the
observed distribution of rotation amplitudes originates in the tested random
walk process is rejected.
For cell numbers lower than 50 the hypotheses that the observed EVPA rotations
are produced by the siQu, the orQUc, or the orQUd process are rejected at
$1\sigma$,  $2\sigma$, and $3\sigma$ significance, respectively, regardless of
the cell variation rate.
In \cref{fig:p3cprob} we have shown that cells fewer than this are needed to
produce the high mean polarization fraction for instance during period~IIIc.
Additionally, cell variation rates $>5$ cells per day are preferred to produce
the observed variability in the polarization fraction.
At these variation rates the hypothesis that the observed EVPA rotations
are produced by one of the tested processes is rejected at least at $2\sigma$
significance.
Therefore, we conclude that the distribution of the rotation amplitudes
identified in the entire EVPA curve cannot be produced by the tested random
walk process with a fixed number of cells.

%___NEW_SECTION________________________________________________________________
\section{Discussion: Two processes?}
\label{sec:discussion}

% Summary of results:
EVPA rotations of opposite direction in \object{3C 279} have been reported
before in the literature \citep{2008A&A...492..389L,2010Natur.463..919A}.
In the data presented here, we observe several changes of the rotation
direction with two large rotations during the periods~IIa and IIIc.
These two periods show significantly different variability in the optical flux
and optical polarization.
During the period~IIa the flux and the polarization fraction are lower and less
variable, whereas the EVPA is more erratic than during the period~IIIc.
Testing the polarization data of both periods against random walk models makes
the difference even more evident.
We have shown that the erratic behaviour of the EVPA during the period~IIa has
the characteristics of a stochastic process.
Although the probability of $\lesssim 3\,\%$ is not high, the discussed models
are capable of producing the observed polarization variability.
Most likely around 170 or 220 cells (depending on the process) are necessary to
produce the observed behaviour.
On the other hand, the high polarization fraction during the period~IIIc
requires significantly fewer cells $\sim 30$, implying that this polarized
emission is produced in a much smaller region.
We have shown that the long and smooth rotation of the EVPA during this period
is highly unlikely to originate in a stochastic process and we could not
reproduce the observed variability in the polarization fraction and angle with
any of the tested stochastic processes.
A deterministic origin of this EVPA rotation is furthermore supported by
the exact contemporaneity with a strong $\gamma$-ray flaring period.

% Conclusion: two processes:
We come to the conclusion that we likely observe two different
processes responsible for the polarization variability during the
low-brightness state and the flaring state of \object{3C 279}.  During the
low-brightness state the polarization is consistent with a stochastic
process. We have argued that the observed variability is unlikely to
be an observational artefact owing to low signal-to-noise data and
underestimated uncertainties, but it is source-intrinsic. This
variability is consistent with a stochastic process as implemented by
the siQU, orQUc, and orQUd process, modelling a general turbulent flow
(siQU), a disturbance passing through a turbulent medium or,
vice versa, a turbulent medium passing through a local disturbance
(orQUc, orQUd). These models are not distinguishable, except from
requiring different model parameters to produce the observed
polarization. We suggest that the polarization during the
low-brightness state is affected by the turbulent flow with a
relatively large number of emission cells having individual magnetic
field orientations, thus producing a relatively low mean polarization
fraction. We suggest that this stochastic variability of the
polarization is always present, but only dominant during the
low-brightness state.

In the data we find two phases which might be transitions from a deterministic
to a stochastic process.
In period~I the initial high activity state in optical and $\gamma$-rays 
dissolves.
With the fading flares the polarization fraction drops and the EVPA variability
becomes far more erratic (c.f. \cref{fig:3C279polopt}).
Again, throughout period~IVb and towards the end of IVc, when the flux density
is decreasing, the EVPA variability estimator rises.
These phases could imply that with the flaring emission fading the turbulent
background becomes more dominant again.

%___NEW_SECTION________________________________________________________________
\subsection{Deterministic process during period~IIIc}
\label{sec:deterministic}

% Period IIIc: deterministic models:
During the flaring state the polarization fraction is on average
higher than during the low-brightness state, indicating that the polarized flux
is dominated by a smaller emission region.
The polarization variability cannot be explained by the tested stochastic
models.
Although we only tested very simplistic random walk processes, this result
challenges more sophisticated stochastic models such as the turbulent
extreme multi-zone model (TEMZ) \citep{2014ApJ...780...87M}.
In contrast to stochastic models, deterministic models are expected to produce
smooth EVPA rotations, some of them following distinct patterns.
In the following we discuss several models presented in the literature.

% Shock compression of tangled field (Laing, 1980)
% and two-component model (Holmes, 1984):
\Citet{1980MNRAS.193..439L} discusses the shock compression of a
tangled magnetic field into an apparently ordered one to
explain a high polarization fraction despite a generally random field
structure. In principle, this process can produce an EVPA swing if the
initially tangled magnetic field produces a low net polarization which
is oriented differently than compressed, apparently ordered field.
But this effect can only produce swings up to $90\,^\circ$ at maximum.
The more generic case of two superposed, evolving emission features
discussed by \citet{1984MNRAS.211..497H} is also limited to swings
$<90\,^\circ$. The observed $\sim360\,^\circ$~rotation excludes these
models.

% Bending jet model:
Two purely geometric models are based on curved trajectories of the emission
region.
In the first model, \citet{2010IJMPD..19..701N} assumes an axially symmetric
magnetic field and a global bend of the flow.
Relativistic aberration can drastically change the observed EVPA of the
emission region going through even a small bend, if also the viewing angle is
small.
Yet, the EVPA rotations in this model are limited to $<180\,^\circ$ for a
simple bend on a plane \citep{2010IJMPD..19..701N}.
\citet{2010Natur.463..919A} used this model to explain the apparent
$\sim208\,^\circ$~EVPA swing in \object{3C 279} within the 20~days after
$\mathrm{JD}2454880$.
We have shown, though, that this apparent rotation was an artefact of the
sparse sampling and that additional data does not support the smooth, long
trend initially reported.
\citet{2014A&A...567A..41A} observed the period~IIIc rotation only partially
with an amplitude of $\sim140\,^\circ$ and used the same bending jet model to
explain this event.
The full $\sim360\,^\circ$~rotation cannot be explained by a single flaring
region on a slightly bent trajectory.
This rotation would either require a helical jet structure or at least two
successive emission regions passing through the bend.
Although we observe several sub-flares in the $R$-band light curve during the
EVPA rotation, these consecutive flaring regions would likely need
significant fine-tuning in the time separation of the individual events to
produce a continuous $360\,^\circ$~rotation.

% Helical trajectory in helical magnetic field:
The second geometric model describes an emission feature smaller than the cross
section of the jet on a helical trajectory traversing through a helical
magnetic field \citep{1988A&A...190L...8K,2008Natur.452..966M}.
On its path the feature highlights different parts of the magnetic field
structure producing a gradually changing EVPA. In contrast to the previous
models this one is not restricted by an upper limit of the rotation amplitude,
but can naturally produce in principle arbitrarily long rotations.
The deviations from a constant rotation rate can originate, e.g. in changes
in the flow speed, varying light-travel time delays along the helical path, or
the superposition of the emission feature and a constantly (or randomly)
polarized background \citep[e.g.][]{2010ApJ...710L.126M,2013ApJ...768...40L}.

% Zhang et al. model:
\citet{2014ApJ...789...66Z,2015ApJ...804...58Z} suggest that EVPA swings can
occur owing to light travel time effects when a relativistically moving plasma
pervading a helical magnetic field encounters disturbance such as a shock.
In contrast to previously discussed models this one naturally explains
contemporaneous EVPA rotations and flaring activity throughout the entire
spectrum from microwaves up to $\gamma$-ray emission, which could explain the 
contemporaneity of the EVPA rotation and the $\gamma$-ray flaring during 
period~IIIc (\cref{fig:3C279polopt}).
However, in the model by \citet{2014ApJ...789...66Z} a single event can only
produce EVPA rotations up $180\,^\circ$.
To explain the observed rotation in the period~IIIc, at least two successive
flaring events would be needed.
There are two sharp drops observed in the polarization fraction during the EVPA
rotation in the period~IIIc, which qualitatively fits the behaviour expected
from the \citet{2014ApJ...789...66Z} model.
However, detailed modelling will be needed to test whether the non-stochastic
$\sim360\,^\circ$~EVPA rotation during the period~IIIc can be produced by the
non-axisymmetric helical motion model or by the axisymmetric model of a shock
in a helical magnetic field.

% Critical challenge to deterministic processes: changes of rotation direction:
The most critical challenge to deterministic models producing large EVPA
rotations (${\geq} 180\,^\circ$) are the multiple inversions of the rotation
direction observed throughout the entire polarization data.
Different directions in the non-flaring and in the flaring state can be
explained with the two-processes interpretation, as a stochastic process during
the non-flaring state produces rotations in both directions equally likely.
But we observe inversions of the rotation direction even within period~IIIc.
Deterministic models that are capable of producing two-directional EVPA swings
such as the two-component model \citep{1984MNRAS.211..497H} and some of the
models presented in \citet{2014ApJ...789...66Z} typically are limited to
amplitudes ${\leq} 90\,^\circ$, producing a rotation followed by a
counter-rotation back to the original orientation;
whereas a single bend in the jet, a helical path of the emission feature in a
helical magnetic field and the models in \citet{2014ApJ...789...66Z} that
produce $180\,^\circ$~rotations are expected to be geometrically bound to a
single EVPA rotation direction.
The superposition of these models with a constantly or randomly polarized
background and the superposition of multiple events overlapping in time could
produce more complex patterns in the EVPA variability.
For example, two emission regions, each on its own may produce a clockwise
rotation of the polarization angle.
Following the two-component model of \citet{1984MNRAS.211..497H}, the
superposition of both regions may temporarily produce a counter-clockwise
rotation of less than $90\,^\circ$, if these regions change their physical
properties as the total intensity or spectral index, for instance when a new
emission feature progressively outshines the previous one.
This superposition could lead to two inversions of the rotation direction as
seen in period~IIIc.
Considering more than one component, of course, drastically increases the model
complexity.

%___NEW_SECTION________________________________________________________________
\subsection{Comparison with recent RoboPol results}
\label{sec:robopol}

\citet{2015MNRAS.453.1669B} investigate the potential connection between
$\gamma$-ray flares and rotations of the optical polarization angle in a
statistical way based on a large sample of blazars monitored with the RoboPol
instrument.
They show that a stochastic process can in principle produce the rotation
amplitudes observed, but it is statistically unlikely that all rotations are
produced by a stochastic process.
This result is consistent with our conclusion that even within the same object
two different processes -- stochastic and deterministic -- may be responsible
for different rotations.
Furthermore, \citet{2015MNRAS.453.1669B} argue that optical EVPA swings and
$\gamma$-ray activity are not necessarily physically connected, but it is
unlikely that none of the observed events are connected.
Particularly, the strongest $\gamma$-ray flares had time lags to EVPA rotations
close to zero.
They conclude there are two processes: one producing strong $\gamma$-ray flares
and contemporaneous rotations of the optical polarization angle; the other
producing moderate $\gamma$-ray flares physically not connected to the
optical polarization activity, which may be coincident happening owing to a
stochastic process.
These results are consistent with ours.
Period~IIIc is an example of a non-deterministic EVPA rotation contemporaneous
with strong $\gamma$-ray flaring activity, whereas the rotation during
period~IIa is probably owing to a stochastic process, potentially not connected
to the moderate $\gamma$-ray variability.
Complementary to \citet{2015MNRAS.453.1669B}, we show that different processes
responsible for the polarization variability can occur in the same object.
Additionally, we show that strong $\gamma$-ray flaring activity can occur
without strong variability in the optical EVPA (period~I) and vice versa
(period~IVa--IVc).

%___NEW_SECTION________________________________________________________________
\subsection{Comparison with \object{BL Lac} and \object{PKS 1510-089}}
\label{sec:comparison}

\object{3C 279}, \object{PKS 1510-089}, and \object{BL Lac} are the most
prominent sources showing rotations of the optical EVPA and contemporaneous, 
strong $\gamma$-ray activity.
The rotations in \object{BL Lac} in 2005 and in \object{PKS 1510-089} in 2009
were explained by a deterministic process, the motion of an emission feature on
a spiral path in a helical magnetic field
\citep{2008Natur.452..966M,2010ApJ...710L.126M}.
The rotations in \object{PKS 1510-089} in 2012, on the other hand, showing
several inversions of the rotation direction, were interpreted as originating
from a stochastic process \citep{2014A&A...569A..46A}.
These events show some similarities but also differences in the progression
of the total optical flux and the polarization fraction during the rotation
of the polarization angle.

The rotations in \object{BL Lac} in 2005 and in \object{PKS 1510-089} in 2009
both start and end with a strong peak in the polarization fraction
\citep{2008Natur.452..966M,2010ApJ...710L.126M}.
The $380\,^\circ$ and $250\,^\circ$~rotations in \object{PKS 1510-089} in 2012
do not exhibit as pronounced peaks, but indicate maxima in the polarization
fraction in the beginning and end of the rotation and a minimum during the
event \citep{2014A&A...569A..46A}.
The rotation in \object{3C 279} during period~IIIc also starts and ends with 
pronounced peaks in the polarization fraction, though in contrast it shows an
increasing trend throughout the event underlying multiple extrema.
This increasing trend of the polarization fraction with a sharp drop at the end
of the rotation is seen in all smooth rotation events, periods~IIIa, IIIc, IVa,
IVb.
In contrast, the erratic rotation in period~IIa does not show a global
trend and no pronounced peak of the polarization fraction at the end of the
rotation.
The beginning of the rotation is not captured by observations.

All discussed rotation events in \object{BL Lac} and \object{PKS 1510-089} and
the period~IIIc rotation in \object{3C 279} coincide with one or multiple peaks
in the optical light curve and $\gamma$-rays, but there is no obviously
consistent shape of the curves.
Periods~IIa, IIIa, and IVa--IVc in the data of \object{3C 279} show that
rotations of the optical polarization angle can happen without coinciding,
strong variability in $\gamma$-rays.

%___NEW_SECTION________________________________________________________________
\section{Summary and conclusions}
\label{sec:conclusions}

In connection to an intensive multi-wavelength campaign, we have accumulated
and combined multiple data sets of $R$-band photometry and optical polarimetry
measurements into well-sampled flux density and polarization curves of the
$\gamma$-ray loud quasar \object{3C 279} covering a period from Nov.~2008 to
July~2012.
These data capture \object{3C 279} in an optical low-brightness state between
Nov.~2009
and Aug. 2010, followed by an optical flaring state, which coincides with an
increased $\gamma$-ray activity \citep{2014A&A...567A..41A}.
We observe strong optical polarization variation and EVPA rotations during both
states.
We have critically discussed the $n\pi$-ambiguity and different solutions
to treat it and we presented a method to estimate the quality of the data given
the time sampling and the observed rotation rates.
We have shown that the sparsely sampled EVPA variability reported by
\citet{2010Natur.463..919A} does not show a continuous, large rotation when
additional data is considered.
Generally, we find EVPA rotations with both a clockwise and a counter-clockwise
sense of rotation.
These multiple changes of the rotation direction observed in the data eliminate
all except perhaps the most geometrically complex models, to explain the entire
polarization variability.

Instead, we have tested whether or not the observed EVPA rotations can be of
stochastic origin using the smoothness of the EVPA curve as a key indicator.
To do this, we introduced the EVPA variation estimator as a quantitative
measure of the curve smoothness.
We simulated three different processes based on random walks in Stokes $Q$ and
$U$ and found that all of them are highly unlikely to produce a smooth
$\sim360\,^\circ$~EVPA rotation as observed during the period~IIIc (the
flaring state), especially coinciding with the high polarization fraction mean
and variation.
We conclude that the tested class of simplistic stochastic processes based on
emission cells that have a random and variable magnetic field orientation
cannot produce the observed polarization variation during the flaring state of
\object{3C 279}.
This result challenges more sophisticated stochastic models such as the 
turbulent extreme multi-zone model (TEMZ) \citep{2014ApJ...780...87M}.

However, in the low-brightness state (period~IIa) the EVPA variation
is much more erratic than during the flaring state and has the characteristics
of random walk processes.
Hence, we find two different states: the low-brightness state exhibits a
comparably low polarization fraction and erratic EVPA variation, possibly
consistent with stochastic variation; the flaring state shows a high
polarization fraction and very smooth EVPA variation with a continuous
counter-clockwise rotation sense.
We interpret this result as two different processes contributing to the
polarization variation in \object{3C 279}.
On the one hand, there probably exists an underlying stochastic variation,
which is visible in the low-brightness, less polarized state.
Any stochastic model naturally explains the frequent changes of the rotation
direction observed during the corresponding period.
On the other hand, during the flaring state a small, highly polarized region of
the optical jet dominates the total and the polarized flux.
The polarization variation in this small region is not produced by the class of
stochastic processes we tested.
Yet, deterministic models are challenged by multiple changes of the rotation
direction observed even during the flaring period.
For period~IIIc we can certainly exclude the bending jet scenario, which had
been suggested by \citet{2014A&A...567A..41A}, who only observed a part of this
EVPA rotation, which significantly exceeds $180\,^\circ$.

We have tested three simplistic random walk processes against polarization
data.
Whereas these toy models neglect various physical effects, we have demonstrated
ways to compare also more sophisticated models statistically with polarization
data.

%ACKNOWLEDGEMENTS%%%%%%%%%%%%%%%%%%%%%%%%%%%%%%%%%%%%%%%%%%%%%%%%%%%%%%%%%%%%%%

\begin{acknowledgements}
  S.K. was supported for this research through a stipend from the International
  Max Planck Research School (IMPRS) for Astronomy and Astrophysics at the Max
  Planck Institute for Radio Astronomy in cooperation with the Universities of
  Bonn and Cologne.
  T.S. was partly supported by the Academy of Finland project 274477.
  The research at Boston University was partly funded by NASA Fermi GI grant
  NNX11AQ03G.
  K.V.S. is partly supported by the Russian Foundation for Basic Research
  grants 13-02-12103 and 14-02-31789.
  N.G.B. was supported by the RFBR grant 12-02-01237a.
  E.B., M.S. and D.H. thank financial support from UNAM DGAPA-PAPIIT through
  grant IN116211-3.
  IA acknowledges support by a Ramon y Cajal grant of the Spanish Ministry of
  Economy and Competitiveness (MINECO).
  The research at the IAA-CSIC and the MAPCAT program are supported by the
  Spanish Ministry of Economy and Competitiveness and the Regional Government
  of Andaluc\'ia (Spain) through grants AYA2010-14844, AYA2013-40825-P, and
  P09-FQM-4784.
  The Calar Alto Observatory is jointly operated by the Max-Planck-Institut
  f\"ur Astronomie and the Instituto de Astrof\'isica de Andaluc\'ia-CSIC.
  Data from the Steward Observatory spectropolarimetric monitoring project were
  used.
  This program is supported by Fermi Guest Investigator grants NNX08AW56G,
  NNX09AU10G, NNX12AO93G, and NNX14AQ58G.
  St.Petersburg University team acknowledges support from Russian RFBR
  grant 15-02-00949 and St.Petersburg University research grant 6.38.335.2015.
  The Abastumani team acknowledges financial support of the project
  FR/638/6-320/12 by the Shota Rustaveli National Science Foundation under
  contract 31/77.  
  We acknowledge the photometric observations from the AAVSO International
  Database contributed by observers worldwide and used in this research.
  This paper has made use of up-to-date SMARTS optical/near-infrared light
  curves that are available at \url{http://www.astro.yale.edu/smarts/glast/}
  \cite{2012ApJ...756...13B}.
  We acknowledge the contributions of Y.~Y.~Kovalev to the Quasar Movie
  Project.
  We thank the anonymous referee for a constructive review that improved
  this paper.
\end{acknowledgements}

%BIBLIOGRAPHY%%%%%%%%%%%%%%%%%%%%%%%%%%%%%%%%%%%%%%%%%%%%%%%%%%%%%%%%%%%%%%%%%%
\bibliographystyle{aa}
\bibliography{references}

%APPENDIX%%%%%%%%%%%%%%%%%%%%%%%%%%%%%%%%%%%%%%%%%%%%%%%%%%%%%%%%%%%%%%%%%%%%%%
\appendix

%_NEW_SECTION__________________________________________________________________
\section{EVPA ambiguity, consistency and reliability}
\label{app:EVPAadjustment}

In the attempt to solve the $n\pi$-ambiguity of the polarization
angle, we generally adjust EVPA curves applying \cref{eq:adjustEVPA1},
in the following we call this method E1. Here, we introduce a second
method (E2) to assess the sampling quality and to test the reliability
of solving the phase wraps. E2 determines the median $\left\{ \cdot
\right\}$ of the $N_\mathrm{ref}$ previous data points as a reference
for an EVPA data point $\chi_i$:

\begin{align}
  \chi_{i,\mathrm{ref}} = \left\{ [\chi_{i-1-N_\mathrm{ref}},\dots, \chi_{i-1}]
  \right\}.
\end{align}

The data point $\chi_i$ is then shifted according to
\cref{eq:adjustEVPA1}, where $\chi_{i,\mathrm{ref}}$ replaces
$\chi_{i-1}$.  For $N_\mathrm{ref} {=} 1$ method~E2 is identical to
E1.  For $N_\mathrm{ref} {>} 1$ method~E2 considers a longer time
frame than E1 as reference for adjusting each data point. It is
expected to be less susceptible to outliers but less reliable for
sparsely sampled rotations.

The measured EVPA curve $\chi_\mathrm{obs}$ is the
$180\,^\circ$-modulo of the intrinsic curve $\chi_\mathrm{intr}$.
EVPA adjustment methods E1 and E2 aim to reconstruct the intrinsic
EVPA variations from the sampled EVPA curve. We call the probability
that the shapes of the adjusted and intrinsic EVPA curves are
identical $P(\chi_\mathrm{adj} {=} \chi_\mathrm{intr} {\pm} n {\cdot}
180\,^\circ)$ (with $n \in \mathbb{N}$) the reliability of each
method. In the following we test the reliability under various
conditions.

Assuming a constant rotation rate $\dot{\chi}$, constant time sampling
$\Delta t$, and no observational errors the EVPA curve is correctly
reconstructed if the sampled rotation $\Delta \chi = \Delta t \cdot
\dot{\chi} < 180\,^\circ /(N_\mathrm{ref} + 1)$.  Without
observational noise $N_\mathrm{ref} = 1$ is the optimal choice
allowing the sparsest time sampling. For sampled rotations
$\Delta\chi < 60\,^\circ$ methods~E1 ($N_\mathrm{ref} {=} 1$) and E2
($N_\mathrm{ref} {\geq} 2$) may yield the same adjusted EVPA curve. We
call the highest number of reference points $N_\mathrm{cons}$, for
which all adjusted curves with $N_\mathrm{ref} \leq N_\mathrm{cons}$
are identical, the \emph{consistency level}:

\begin{align}
  N_\mathrm{const} = \max \{N \in \mathbb{N}_{\geq1} :  \chi_{\mathrm{adj},
  (N_\mathrm{ref}=1)} = \chi_{\mathrm{adj}, (N_\mathrm{ref}=i)} \
  \forall\  i \leq N\}.
  \label{eq:consistencylevel}
\end{align}

Assuming minimal variation between measured data points, the
consistency level allows one to estimate the reliability of the EVPA
curve reconstruction.

\begin{figure}
  \centering
  \includegraphics[width=\columnwidth]{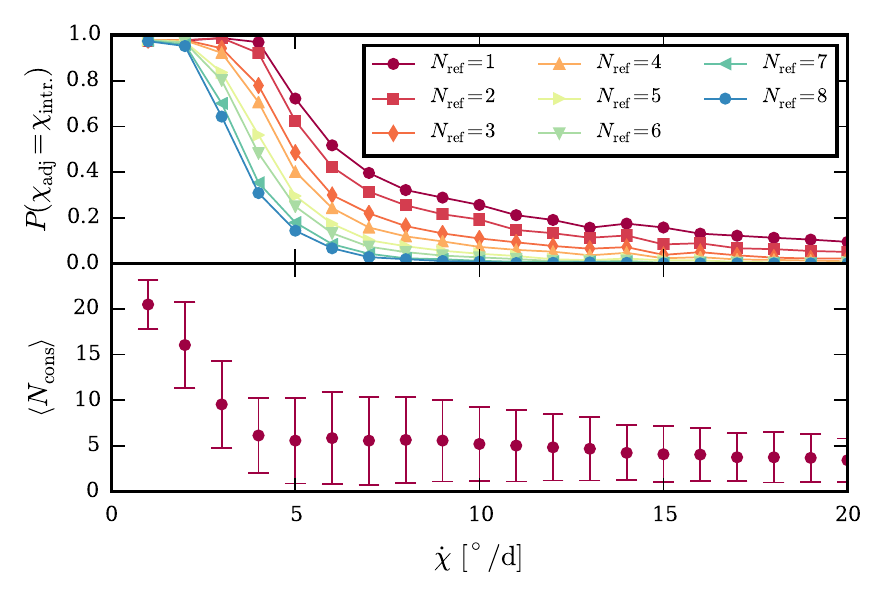}
  \caption{Probability of the correct EVPA reconstruction using different     
           numbers of reference points $N_\mathrm{ref}$ (upper panel) and
           average consistency levels (lower panel) over different rotation
           rates $\mathrm{d}\chi/\mathrm{d}t$.}
  \label{fig:adjustRotation}
\end{figure}

We determine the expected reliability of the EVPA adjustment (E1 and
E2) and the expected consistency level for various intrinsic rotation
rates given the time sampling and observational errors of our combined
polarization data.  We simulate linearly increasing EVPA curves up to
a total amplitude of $360\,^\circ$ with random time sampling and
random errors as in the random walk models
(c.f.~\cref{sec:randomwalks}) with model parameters given in
\cref{tab:secmodelpars}, row 'total'.  The upper panel of
\cref{fig:adjustRotation} shows the probability of correctly
reconstructing the intrinsic EVPA variation for various intrinsic
rotation rates and different numbers of reference points.  Method~E1
($N_\mathrm{ref} {=} 1$) is the most reliable for rotation rates
$\dot{\chi} \geq 3\,^\circ/\mathrm{d}$, increasing $N_\mathrm{ref}$
reduces the reliability.  For intrinsic rotation rates $\dot{\chi}
\leq 3\,^\circ/\mathrm{d}$ the reliability is $\sim 98\,\%$.  In the
lower panel of \cref{fig:adjustRotation} we show the expectation value
and standard deviation of the consistency level for different
intrinsic rotation rates.  High consistency rates
$N_\mathrm{cons}\gtrsim 10$ indicate low intrinsic variation
$\dot{\chi} \leq 3\,^\circ/\mathrm{d}$ and a high probability $\sim
98\,\%$ of correct curve reconstruction.

Given the time sampling and typical uncertainties of our polarization
data, method~E2 does not improve the reliable reconstruction of the
intrinsic EVPA variability. Nevertheless, using both methods E1 and E2
and the consistency level allows us to estimate the quality of the time
sampling and the reliability of the reconstructed EVPA curve.

%_NEW_SECTION__________________________________________________________________
\section{Variation estimator biases}
\label{app:varestbias}

% Error bias:
We have defined a quantitative measure of the smoothness of the EVPA
curve in \cref{sec:variationestimator}. This variation estimator
measures the average, short time-scale (shorter than the analysed data
total time), erratic rotation rate of the EVPA corrected for an
assumed, underlying, secular trend at the time-scale of the analysed
data total time. The measured variation estimator is increased (a) by
measurement errors, introducing additional variation, and (b) by
curvature of the underlying smooth variation, i.e. a non-constant
trend.

Intrinsic EVPA variation and measurement errors independently add to
the variation estimator. The de-biased variation estimator is given
by:

\begin{align}
  s_\mathrm{debiased} = \sqrt{s_\mathrm{obs}^2 - s_\mathrm{bias}^2}
  \text{\ \ for\ \ } s_\mathrm{obs} > s_\mathrm{bias}.
  \label{eq:varestdebias}
\end{align}

For constant measurement uncertainties $\sigma_\chi$ and even time
sampling $\Delta t$, the error bias of the variation estimator is
$s_\mathrm{bias} \propto \sigma_\chi / \Delta t$.
\Cref{eq:varestdebias} also holds for variable $\sigma_\chi$, but then
we need to estimate $s_\mathrm{bias}$ through simulations, which is
how the values in \cref{tab:polarization} were estimated.  The
uncertainty of the de-biased variation estimator is equal to the
uncertainty of the error bias.

\begin{figure}
  \centering
  \includegraphics[width=\columnwidth]{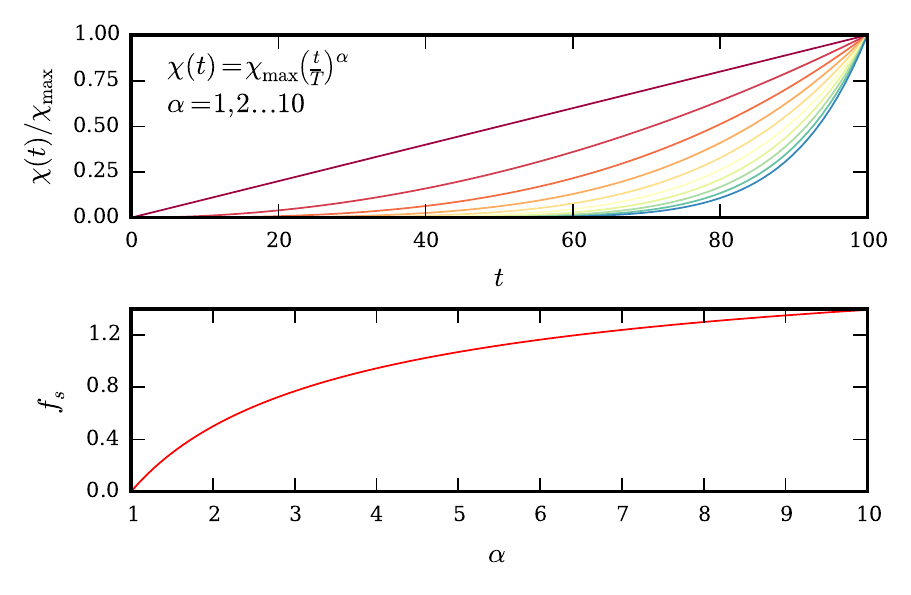}
  \caption{The upper panel shows exemplary power law curves with increasing 
           curvature, i.e. power law index $\alpha$ from $1$ (red) to $10$
           (blue). The lower panel shows the curvature factor (see text for
           description).}
  \label{fig:varestcurvebias}
\end{figure}

% Curvature bias:
When calculating the variation estimator $s$, the local derivative is corrected
for (intrinsic) first order rotations,
i.e. constant rotation rates, by subtracting a trend $\bar{\chi}_t$.
For higher order (i.e. curved) intrinsic rotation curves the variation
estimator will be biased depending on the curvature.  We test this
dependency by simulating EVPA curves (without errors) that follow a
power law over time

\begin{align}
  \chi(t) = \chi_\mathrm{max} \left(\frac{t}{T}\right)^\alpha,
\end{align}

with $\chi_\mathrm{max}$ the EVPA at time $T$ equal to the total
rotation amplitude during the total time interval $T$ and with $\alpha
\geq 1$ the power law index.  Example curves for $\alpha = 1, 2,
\cdots, 10$ are shown in the upper panel of
\cref{fig:varestcurvebias}.

The trend $\bar{\chi}_t$ estimates the total rotation rate given by
$\frac{\Delta\chi}{\Delta t} = \frac{\chi_\mathrm{max}}{T}$ with an
accuracy of $1\,\%$.  The curvature biased variation estimator is
given by the curvature factor $f_s$ (plotted in the lower panel of
\cref{fig:varestcurvebias}) and the total rotation rate
$\frac{\Delta\chi}{\Delta t}$ estimated by the trend $\bar{\chi}_t$:

\begin{align}
  s_\mathrm{curve} = f_s \cdot \frac{\Delta\chi}{\Delta t} \approx f_s \cdot 
\bar{\chi}_t.
  \label{eq:varestcurvebias}
\end{align}

\Cref{eq:varestcurvebias} with the simulation result shown in
\cref{fig:varestcurvebias} allows us to estimate the impact of a curved,
underlying, smooth variation of the EVPA onto the variation estimator.
Correcting the variation estimator for higher order variation would
require fitting the EVPA curve and subtracting the fit from the
local derivatives. To avoid a priori fitting an arbitrary
function to the data we stick to the criterion that a linear trend is
considered smooth variation and any deviation from that linear trend
contributes to the variation estimator.

%_NEW_SECTION__________________________________________________________________
\section{EVPA rotation identification and distribution}
\label{app:rotident}

We define an algorithm to automatically identify EVPA rotations in a
polarization curve based on strictly fixed criteria.  We use this
algorithm on polarization curves simulated with our random walk
processes to characterize the EVPA rotation amplitude distribution
expected from a stochastic process.

%_NEW_SECTION__________________________________________________________________
\subsection{EVPA rotation identification algorithm}
\label{app:rotident:identification}

We define the start and end point of an EVPA rotation by a change of
the rotation direction, i.e. a sign change of the derivative of the
EVPA curve.
We do not consider a change of the rotation rate. 
We define the rotation \emph{amplitude} as the difference of the (local)
extrema: $\left| \Delta\chi \right| = \chi_\mathrm{max} - \chi_\mathrm{min}$. 
We call a rotation between two data points $\chi_i$, $\chi_j$
\emph{significant} when the amplitude is larger than the root summed squared
errors of the two data points multiplied with a factor $\varsigma$ that
characterizes the significance level:

\begin{align}
  \left| \Delta\chi_{ij} \right| = \left| \chi_i - \chi_j \right|
  > \varsigma \cdot
  \sqrt{\sigma_{\chi_i}^2 + \sigma_{\chi_j}^2}.
  \label{eq:significance}
\end{align}

The algorithm identifies significant rotations by iterating pointwise
through the data.  The \emph{storage} contains all identified,
significant EVPA rotations.
The \emph{current rotation} contains all data points between the end
of the last significant rotation and the data point previous to the
current one.
The rotation amplitude, direction and significance characterize the current
rotation.
They are calculated from the minimum and maximum value and the corresponding
uncertainties.
The data point of the current iteration step and the previous one make the
\emph{new rotation}.

\definecolor{Gray}{gray}{0.9}
\begin{table*}
  \caption{Description of the occurring states and the corresponding operations 
           for the EVPA rotation identification.}
  \label{tab:rotidentalgorithm}
  \begin{tabular*}{\linewidth}{@{\extracolsep{\fill} } l p{0.08\linewidth} 
p{0.08\linewidth} p{0.08\linewidth} p{0.08\linewidth} p{0.08\linewidth} | 
p{0.12\linewidth} p{0.12\linewidth} p{0.12\linewidth}}
    \toprule
    & \multicolumn{5}{c |}{\textbf{State}}   & 
\multicolumn{3}{c}{\textbf{Operations}}
    \\\midrule
    &
    \parbox[t]{1.0cm}{Current\\ rotation\\ significance}  &
    \parbox[t]{1.0cm}{New\\ rotation\\ significance}  &
    \parbox[t]{1.0cm}{Insign.\\ rotation\\ becoming\\ significant}  &
    \parbox[t]{1.0cm}{Rotation\\ direction\\ current/\\ new}  &
    \parbox[t]{1.0cm}{Rotation\\ direction\\ stored/\\ current}  &
    \parbox[t]{1.0cm}{Store\\ rotation}  &
    \parbox[t]{1.0cm}{Set\\ current\\ rotation}  &
    \parbox[t]{1.0cm}{Set\\ current\\ rotation\\ significance}
    \\\midrule
    \rowcolor{Gray}
    1.    & Y   & Y   &     & +   &     &                                                & current+new                                        & Y  \\
    2.    & Y   & Y   &     & -   &     & current                                        & new                                                & Y  \\
    \rowcolor{Gray}
    3.    & Y   & N   &     & +   &     &                                                & current+new                                        & Y  \\
    4.    & Y   & N   &     & -   &     & current                                        & new                                                & N  \\
    \rowcolor{Gray}
    5.    & N   & N   & N   &     &     &                                                & current+new                                        & N  \\
    6.    & N   & N   & Y   &     & n/a &                                                & \parbox[t]{1.0cm}{min/max\\ (current)+new}         & Y  \\
    \rowcolor{Gray}
    7.    & N   & N   & Y   &     & +   & delete last entry                              & \parbox[t]{1.0cm}{stored+min/max\\ (current+new)}  & Y  \\
    8.    & N   & N   & Y   &     & -   & \parbox[t]{1.0cm}{stored+min/max\\ (current)}  & \parbox[t]{1.0cm}{min/max\\ (current)+new}         & Y  \\
    \rowcolor{Gray}
    9.    & N   & Y   &     &     & n/a &                                                & \parbox[t]{1.0cm}{min/max\\ (current)+new}         & Y  \\
    10.   & N   & Y   &     & +/- & +/- & delete last entry                              & \parbox[t]{1.0cm}{stored+current\\ +new}           & Y  \\
    \rowcolor{Gray}
    11.   & N   & Y   &     & +/- & -/+ & \parbox[t]{1.0cm}{stored+min/max\\ (current)}  & \parbox[t]{1.0cm}{min/max\\ (current)+new}         & Y  \\
    \bottomrule
  \end{tabular*}
\end{table*}

Based on the direction and significance of the new, current, and last stored
rotation we define eleven states and corresponding operations in
\cref{tab:rotidentalgorithm}. 
Five quantities define each state: the significance of the current and the new
rotation (Y/N, col.~2 and~3), whether the current rotation changes its
direction or not (+/-, col.~5), whether the current rotation becomes
significant by adding the new, current data point(Y/N, col.~4), and how the
newly significant rotation is oriented regarding to the previously stored
rotation (+/-, col.~6).
States~6 and~9 may only occur once when no significant rotation has been found
yet and the rotation direction cannot be compared to a previous rotation (n/a).
This scheme only identifies local maxima and minima, which are significant at a
predefined level.
Each pair of adjacent identified extrema marks a continuous rotation, which may
contain several local, insignificant maxima and minima.

%_NEW_SECTION__________________________________________________________________
\subsection{Random walk EVPA rotation distribution}
\label{app:rotident:distribution}

\begin{figure}
  \centering
  \includegraphics[width=\columnwidth]{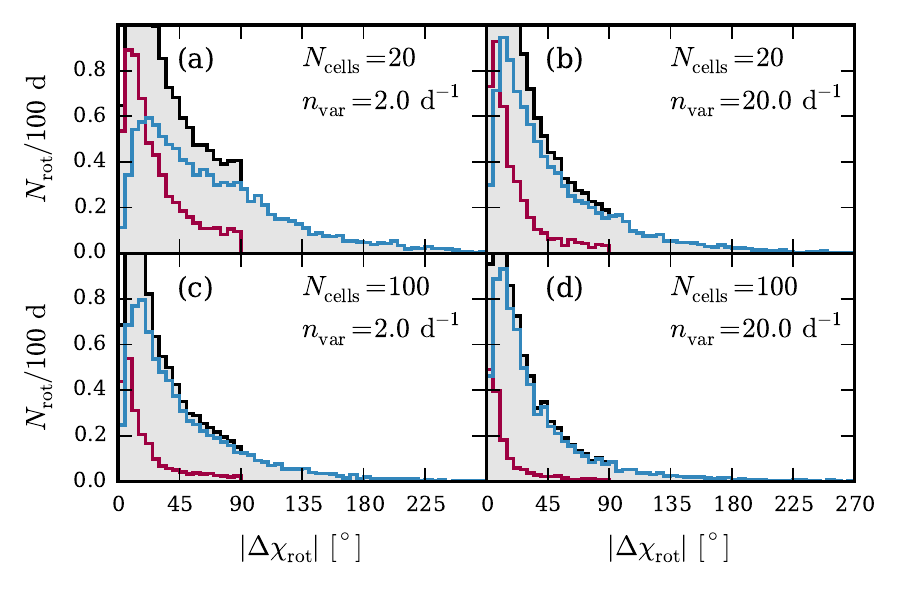}
  \caption{Distribution of the absolute amplitudes   
           $\left|\Delta\chi_\mathrm{rot}\right|$ of rotations  identified at   
           $3\sigma$-level in siQU model simulations with different cell
           numbers and cell variation rates. Red line: distribution of
           two-data-point rotations, blue line: distribution of
           multiple-data-point rotations, grey filled region: distribution of
           all rotations.}
  \label{fig:rotdistrw}
\end{figure}

We run random walk simulations for various cell numbers and cell variation
rates with a total time $T{=}100\,000\,\mathrm{d}$ and the other parameters
chosen as in \cref{tab:secmodelpars}, period~``total''.
We adjust the EVPA curve with \cref{eq:adjustEVPA1} and identify significant
rotations at a significance level $\varsigma{=}3$.
In \cref{fig:rotdistrw} we show four examples of EVPA rotation amplitude
distributions based on the siQU random walk process with different numbers of
cells and cell variation rates.
The grey filled region displays the full distribution.
Two populations of identified rotations contribute to this distribution:
rotations consisting of two data points only, $N_\mathrm{dp}{=}2$, and
rotations containing more than two data points, $N_\mathrm{dp}{>}2$.
The rotation amplitude distributions of the populations are shown as red and
blue curves, respectively.

Both distributions peak at a certain value, show a sharp decline towards
smaller amplitudes and a long tail towards larger amplitudes.
However, because of the $n\pi$-ambiguity the two-data-point-population
distribution is cut off at $\sim 90\,^\circ$.
Slightly larger rotation amplitudes than $90\,^\circ$ are possible owing to 
EVPA errors.
The shortest simulation time step and the limits of the tested parameter space
(cell number and cell variation rate) affect the low-amplitude tail of the
distributions.

When the cell variation rate increases (relative to the number of
cells) the EVPA variation becomes more erratic and large, continuous rotations
become less likely.
The distribution shifts towards smaller rotation amplitudes.
When the number of cells is reduced the EVPA variation becomes more
erratic and the EVPA uncertainty is lower as the average polarization fraction
is increased.
As a result, two-data-point rotations are more likely to be significant and the
number of these rotations increases, whereas the number of small rotation
consisting of multiple data points decreases.

Though not shown here, the significance level $\varsigma$ used in the
rotation identification strongly affects the number of identified
rotations and the rotation amplitude distribution.  For larger
$\varsigma$ two-data-point-rotations are less likely to be significant
and therefore the number of identified rotation tends to be lower, and
rotation amplitudes tend to be larger compared to a lower choice of
$\varsigma$.

\end{document}